\documentclass[vecphys]{svmult}

\usepackage{makeidx}         
\usepackage{graphicx}        
\usepackage{multicol}        
\usepackage{cite}            
\usepackage[bottom]{footmisc}
\usepackage{amsmath,amssymb}
\smartqed

\makeindex

\newtheorem{thm}{Theorem}
\newtheorem{lem}{Lemma}
\newtheorem{cor}{Corollary}
\newtheorem{prop}{Proposition}
\newtheorem{As}{Assumption}

\newenvironment{pr}{\vspace{1mm}\noindent\textbf{Proof.}}
                   {\vspace{-5mm}\begin{flushright}$\Box$\end{flushright}}

\newenvironment{pro}[1]{\vspace{1mm}\noindent\textbf{Proof of #1.}}
                   {\vspace{-5mm}\begin{flushright}$\Box$\end{flushright}}

\newenvironment{defin}{\vspace{1mm}\noindent\textbf{Definition.}}{\vspace{1mm}}

\newcommand{\FT}[1]{{#1}^*}

\def\NN{\mathbb{N}}

\def\RR{\mathbb{R}}
\def\CC{\mathbb{C}}
\def\PP{\mathbb{P}}

\def\1{\mathbb{1}}
\def\EE{\mathbb{E}}
\def\EEW{\widetilde{\mathbb E}}

\def\FF{\mathcal{F}}

\def\1{\mathbf{1}}

\def\BEN{\begin{enumerate}}  \def\BI{\begin{itemize}}
\def\EEN{\end{enumerate}}   \def\EI{\end{itemize}} 
   
  \def\no{\noindent}

\def\nn{\nonumber}
\def\beq{\begin{eqnarray}} \def\eeq{\end{eqnarray}}

\def\al*#1{\begin{align*}#1\end{align*}}

\def\ga*#1{\begin{gather*}#1\end{gather*}}

\def\alat*#1#2{\begin{alignat*}{#1}#2\end{alignat*}}
\def\bea{\begin{eqnarray*}}
\def\eea{\end{eqnarray*}}
\def\ml*#1{\begin{multline*}#1\end{multline*}}

\def\mbb{mathbb} \def\mbf{\mathbf} \def\mrm{\mathrm}
 \def\unl{\underline} \def\ovl{\overline}

\def\lb{\ell}
\def\ub{\upsilon}

\def\cu{\mathtt{i}}

\def\mbb{\mathbb}
\def\mc{\mathcal}

\def\le{\left} \def\ri{\right} \def\i{\infty}

\def\te#1{\mathrm{e}^{#1}}  
\def\td{\mathrm{d}}
 
\def\I{\int} 

\def\WH{\widehat} \def\WT{\widetilde}

\def\a{\alpha} \def\b{\beta}
  \def\d{\delta}   

\def\e{\epsilon}  \def\l{\lambda}  
  \def\nn{\nonumber}   \def\s{\sigma}
\def\t{\tau}     
 \def\w{\omega} \def\q{\qquad} 
 \def\G{\Gamma}

\newcommand{\exit}{{\mbox{\, \vspace{3mm}}} \hfill\mbox{$\square$}}

\begin{document}

\title{Exotic derivatives under stochastic volatility models with jumps}

\author{Aleksandar Mijatovi\'{c} \and Martijn Pistorius}
\institute{Department of Statistics, University of Warwick,
\texttt{a.mijatovic@warwick.ac.uk}
\and Department of Mathematics, Imperial College London,
\texttt{m.pistorius@imperial.ac.uk}}

\maketitle

\begin{abstract}
In equity and foreign exchange markets the risk-neutral dynamics of the underlying
asset are commonly represented by stochastic volatility models with jumps.
In this paper we consider a dense subclass of such models
and develop  analytically tractable formulae for the prices
of a range of first-generation exotic derivatives.
We provide closed form formulae for the Fourier transforms of vanilla 
and forward starting option prices as well as a formula for
the slope of the implied volatility smile for large strikes. 
A simple explicit approximation formula for the variance 
swap price is given. The prices of volatility swaps and other volatility derivatives 
are given as a one-dimensional integral of an explicit function. 
Analytically tractable formulae for the Laplace transform (in maturity) 
of the double-no-touch options and
the Fourier-Laplace transform 
(in strike and maturity) of the double knock-out call and put options are 
obtained. The proof of the latter formulae is based on extended
matrix Wiener-Hopf factorisation results.
We also provide convergence results.
\\
\noindent \keywords{Double-barrier options, volatility surface, volatility derivatives, 
forward starting options, stochastic volatility models with jumps, fluid embedding, complex matrix 
Wiener-Hopf factorisation}
\end{abstract}


\section{Introduction}
\label{sec:Intro}
A key step
in the valuation and hedging of exotic derivatives
in financial markets is to  decompose these in terms of simpler securities,
e.g. vanilla options, which trade in larger volumes, are generally very liquid
and therefore have a well defined price.  Such a decomposition is often achieved in two steps. 
First a model for the underlying asset under a risk neutral measure is 
calibrated to the implied volatility surface. In this step the current state 
of the market, as described by the prices of vanilla derivatives, is 
expressed in terms of the parameter values of the model. In other words the 
chosen model is used to impose a structure on the option prices. 
The second step consists of pricing 
the exotic derivative of interest in the calibrated model. 

It is well known that in equity and foreign exchange markets stochastic
volatility models with jumps can be used to accomplish the first step 
described above (see e.g.~\cite{Gatheral},~\cite{Lipton_Book}).
In the present paper 
we consider forward starting vanilla options, volatility derivatives and barrier options, which are among 
the most widely traded exotic derivatives in
the equity and foreign exchange markets. 

The desired properties of the model in each of the two steps described above
place diametrically opposite restrictions on the 
choice of modelling framework. 
This is because in the calibration step
one requires a flexible stochastic process that can describe well the current state of 
the vanilla market (i.e. can calibrate accurately to the observed implied
volatility surface), 
while such flexibility can be a source of problems in the 
second step, where one needs to compute expectations of path-dependent functionals of the 
process. A more rigid modelling framework with, say, continuous trajectories and 
some distributional properties (e.g. independence of increments) could yield  the 
structure of the process needed to establish efficient 
pricing algorithms  for exotic derivatives.

We investigate two families of stochastic volatility models with jumps: 
the time-changed exponential L\'{e}vy models and 
the stochastic volatility models driven by L\'{e}vy processes, where the volatility process is independent of 
the L\'{e}vy driver.
In these two families of  models, the pricing of European derivatives is well understood 
and efficient calibration methods have been developed 
(see for example~\cite{CarrMadanGemanYor_StochVolLevy}
and~\cite{Lipton_Book}), i.e. the first step of the two-stage procedure
outlined above. However
once the model  
is calibrated, the problem of pricing 
the first-generation exotic derivatives (e.g. barrier options) is quite 
involved. The law of the first-exit time from a bounded interval in stochastic volatility
models with jumps, for instance, 
is not usually available in analytically tractable form. 
Because of the lack of 
structural properties 
that can be exploited to find the laws of the path-dependent functionals
of interest, one would typically need to resort to Monte Carlo
methods for the pricing of such derivatives in this setting. 
It is well-known  that
these methods are time-consuming and yield 
unstable results, especially when used to calculate the sensitivities
of derivative securities. 
The method proposed in this paper to calculate the prices of such contracts consists of two steps:
(i) a Markov chain approximation of the volatility process  and (ii) 
an analytically tractable solution of the value function of the contract of interest in the approximating model. 
We provide proofs for the convergence of option prices under this approximation, and derive explicit 
expressions of Laplace and/or Fourier transforms of the value functions under the approximating model.

The approximating class of stochastic volatility processes with jumps considered in this paper retains 
the structural properties required for the semi-analytic pricing (i.e. up to an integral transform) of 
forward starting options, volatility derivatives and barrier options.
In the case of double-barrier option prices we will show that the process considered here
admits 
explicit formulae for the Laplace/Fourier transforms
in terms of the solutions 
of certain quadratic matrix equations. 
The main mathematical contribution of the present paper,
which underpins the derivation of these closed form formulae, is the proof of the existence and uniqueness
of the matrix Wiener-Hopf factorisation of a class of complex valued matrices
related to the approximating model
(see Theorem~\ref{thm:WienerHopf}). These results extend those of \cite{Pistorius_American}
where the corresponding results for the real-valued case 
are established. 
In the context of noisy fluid flow models 
the matrix Wiener-Hopf factorisation 
for the case of a regime-switching Brownian motion
is studied by \cite{Asmussenfluid,Rogersfluid}.

It should be noted that the matrix Wiener-Hopf factorisation results 
developed in this paper
can also be applied to the pricing of American call and put options 
in model~\eqref{eq:Process} as follows.
First we apply the main result in~\cite{Pistorius_American} 
to obtain the price of
the perpetual American call or put and then, via the randomization 
algorithm introduced in~\cite{Carr_rand},
find the actual price of the option.

Related Markov chain mixture models, which are special 
cases of the model considered in this paper,
have been studied before in the mathematical finance literature. 
In~\cite{Guo_Zhang04} and~\cite{Jobert_Rogers06}
explicit formulae were derived for the price of a perpetual American put option
under a regime-switching Brownian motion model.
The same process was used in~\cite{GDG_Rogers06}
to model stochastic dividend rates where 
the problem of the pricing of barrier options on equity was considered.
Finite maturity American put options 
were considered in~\cite{Elliott}
under a regime-switching Brownian motion model.
More generally 
in~\cite{BoyLev_Americanrs},~\cite{BoyLev_Americansi}
numerical algorithms were developed in the case of 
regime-switching L\'evy processes.
Furthermore extensive work has been done on derivative pricing under stochastic 
volatility models with and without jumps (see the standard references~\cite{Lipton_Book},
\cite{Gatheral}, \cite{ContTankov} and~\cite{TheBook}).

The remainder of the paper is organized as follows. 
In Section~\ref{sec:PhaseType}
we state and prove the properties of continuous-time
Markov chains and phase-type distributions
that are needed to define the class of stochastic volatility
models with jumps studied in this paper.
In Section~\ref{sec:Model}
we give a precise definition of this class 
of models and describe an explicit construction of 
an approximating sequence of models, based on 
Markov chains, which converges to the  given stochastic
volatility model with jumps.
The models are set against the backdrop of  a foreign exchange market 
which allows us to include 
naturally the stochastic foreign and domestic discount factors. 
In Section~\ref{sec:EuropVol} we provide explicit formulae for the
Fourier transforms in model~\eqref{eq:Process}
for vanilla and forward starting options. This section also gives an 
approximate explicit formula for the pricing of variance swaps, a
one-dimensional integral representation of the price of a volatility 
swap and formulae for the asymptotic behaviour of the implied
volatility smile for large strikes.
Section~\ref{sec:AD} is devoted to the first-passage times of regime-switching processes.
Section~\ref{sec:DNT_KOC}
discusses the pricing of double-no-touch and double-barrier knock-out
options.  It provides a formula for the single Laplace transform (in maturity)
and the Laplace-Fourier transform (in strike and maturity)
of the double-no-touch 
and the double-barrier knock-out options respectively
in terms of the quantity that can
be obtained from the complex matrix Wiener-Hopf factorisation 
(see Theorem~\ref{thm:Projection}). Section~\ref{sec:FE}
describes the fluid embedding of the model in~\eqref{eq:Process},
which plays a central role in the Wiener-Hopf factorisation.
The key mathematical results of the paper, which allow us to price barrier
options in the setting of stochastic volatility,
are contained in Section~\ref{sec:WH}
where matrix Wiener-Hopf factorisation is defined and
the theorems asserting its uniqueness and existence are stated.
The proofs of these (and other) results are contained in the
appendix.

\section{Markov chains and phase-type distributions}
\label{sec:PhaseType}
\subsection{Finite state Markov chains}

We start by collecting some useful and well-known 
properties of finite state Markov chains
that will be important in the sequel (see e.g.~\cite{Freedman:71}).
For completeness we will also present the proofs.
Throughout the paper we will denote by
$$
M(i,j) = M_{ij} = e_i' M e_j,\qquad m(j) = m_j=m'e_j,\qquad i,j = 1,\ldots, n,
$$
the $ij$th element of an $n\times n$ matrix $M$ and the $j$th element of an $n$-dimensional vector $m$, where the vectors
$e_i$, $i=1,\ldots,n$,
denote the standard basis of
$\CC^n$ and where $'$ means transposition. Throughout the paper 
$I$ will denote an identity matrix of appropriate size, and $\RR_+=[0,\infty)$
the non-negative real line.

\begin{lem}
\label{lem:MC_Semigroup}
Let
$Z$
be a Markov chain on a state space
$E^0:=\{1,\ldots,N_0\}$,
where
$N_0\in\NN$,
and let
$B:E^0\to\CC$
be any function.
If
$Q$
denotes
the generator of
$Z$
and
$\Lambda_B$
is a diagonal matrix of size
$N_0$
with diagonal elements equal to
$B(i)$,
$i=1,\ldots,N_0$,
then it holds that
\begin{eqnarray}
\label{eq:TheFormula}
\EE_i\left[\exp\left(\int_0^tB(Z_s)ds\right)I_{\{Z_t=j\}}\right] & = & 
\exp\left(t(Q+\Lambda_B)\right)(i,j)\qquad \text{ for any } \quad i,j\in E^0,\ t\ge 0,
\end{eqnarray}
where $\EE_i[\cdot] = \EE[\cdot|Z_0=i]$, 
$\PP_i[\cdot] = \PP[\cdot|Z_0=i]$, and $I_{\{\cdot\}}$
is the indicator of the set $\{\cdot\}$.
\end{lem}

\begin{pr}
Let 
$(P_t)_{t\geq0}$
be a family of 
$N_0$-dimensional square matrices with
entries 
$P_t(i,j)$,
$i,j=1,\ldots,N_0$,
given by the left-hand side of~\eqref{eq:TheFormula}.
It is clear that 
$P_0=I$,
where
$I$
is the
$N_0$-dimensional 
identity matrix. 
The Markov property of the chain 
$Z$
yields the Chapman-Kolmogorov equation
$P_{t+s}=P_sP_t=P_tP_s$ 
for all
$ s,t\geq0.$
If we show that the family of matrices 
$(P_t)_{t\geq0}$
satisfies the system of ODEs with constant coefficients
\begin{eqnarray}
\label{eq:DiffEqation}
\frac{\td P_t}{\td t} = (Q+\Lambda_B)P_t,\qquad P_0=I,
\end{eqnarray}
then the lemma will follow, since equation~\eqref{eq:DiffEqation} 
is well-known to have a unique solution given by the right-hand side
of~\eqref{eq:TheFormula}.
The Chapman-Kolmogorov equation implies that 
$ P_{t+h}-P_t= (P_{h}-I)P_t $
and it is therefore enough to show
$\lim_{h\to0}
(P_{h}-I)/h=Q+\Lambda_B$.
In other words we need to prove
\begin{equation}
\label{eq:LimitToShow}
\lim_{h\to0} (P_h(i,j)-I(i,j))/h=
\begin{cases}
Q(i,j) & \text{if}\> i\neq j,\\
Q(i,j) + B(j) & \text{if}\> i=j.
\end{cases}
\end{equation}
The random variables
$B(Z_s)$
are bounded uniformly in 
$s$
and hence the
Taylor expansion of the exponential yields
$$
P_h(i,j) = \EE_i\left[I_{\{Z_h=j\}}\left(1+\int_0^hB(Z_s)\td
s\right)\right]+o(h)\qquad\text{for all}\quad i,j\in\{1,\ldots N_0\}.
$$
It is clear that 
$$\lim_{h\to0}I_{\{Z_h=j\}} \frac{1}{h}\int_0^hB(Z_s)\td s= I_{\{Z_0=j\}} B(Z_0) \quad\PP_i\text{-a.s.}
\qquad\text{for all}\quad i,j\in\{1,\ldots N_0\},$$
since the paths of 
$Z$
are 
$\PP_i$-a.s constant for exponentially distributed amount of time.
The dominated convergence theorem 
and the well-known fact
$\EE_i\left[I_{\{Z_h=j\}} \right]=h(I(i,j)+ Q(i,j))+o(h)$
therefore imply~\eqref{eq:LimitToShow}. 
This concludes the proof.
\end{pr}


We now apply Lemma~\ref{lem:MC_Semigroup}
to esatblish a simple but important property of the
specturm of a discounted generator.

\begin{lem}
\label{lem:spectrum}
Let
$Q$
be a generator of a Markov chain
with
$N_0\in\NN$
states
and let
$D$
be a complex diagonal matrix of dimension
$N_0$.
Then every eigenvalue
$\lambda\in\CC$
of the matrix
$Q-D$
(i.e.
a solution of the equation
$(Q-D)x=\lambda x$
for some non-zero element
$x$
in
$\CC^{N_0}$)
satisfies the inequality
$$\Re(\lambda)\leq -\min\left\{\Re(d_i)\>:\>i=1,\ldots,N_0\right\},$$
where
$d_i=D(i,i)$,
$i=1,\ldots,N_0$,
are diagonal elements of
$D$.
In particular if
$\min\left\{\Re(d_i)\>:\>i=1,\ldots,N_0\right\}>0$,
then the matrix
$Q-D$
is invertible.
Furthermore, the real part of every eigenvalue of $Q$ is non-positive.
\end{lem}

\begin{pr}
Let
$\lambda$
be an eigenvalue of
the matrix
$Q-D$
that corresponds to the eigenvector
$x\in\CC^{N_0}$.
Then
$x$
is also an eigenvector with eigenvalue
$\exp(\lambda)$
of the matrix
$\exp(Q-D)$.
Lemma~\ref{lem:MC_Semigroup}
implies that
if
$Z$
is the chain generated by
$Q$
then the following identity holds
\begin{eqnarray}
\label{eq:MCIdnet}
e_i'\exp(Q-D)x\>\>=\>\>\sum_{j=1}^{N_0}
x_j\EE_i\left[\exp\left(-\int_0^1d_{Z_t}dt\right)I_{\{Z_1=j\}}\right],\quad
i=1,\ldots,N_0,
\end{eqnarray}
where
$e_i$
(resp.
$d_i$)
denotes the
$i$-th
basis vector in
$\CC^{N_0}$
(resp. diagonal element of the
matrix
$D$).

Assume now without loss of generality that
the norm
$\|x\|_\infty:=\max\{|x_i|\>:\>i=1,\ldots,N_0\}$
of the vector
$x$
is one.
Then
identity~\eqref{eq:MCIdnet}
implies
the estimate
$$\exp(\Re(\lambda))=|\exp(\lambda)|=\|\exp(Q-D)x\|_\infty\leq
\exp(-\min\left\{\Re(d_i)\>:\>i=1,\ldots,N_0\right\})
$$
which proves the lemma.
\end{pr}

\begin{lem}
\label{lem:h}
Let $q\in\mathbb C$ be such that $\Re(q)>0$ 
and $M$ a matrix whose eigenvalues all have 
non-positive real part.
Then the matrix $qI - M$
is invertible
and the following formula holds
\begin{eqnarray}
\label{eq:FormulaForh}
\int_0^\infty \te{-qt} \exp(tM)\td t  = 
(qI - M)^{-1}.
\end{eqnarray}
\end{lem}

\begin{pr}
%
The following identity holds for any 
$T\in(0,\infty)$
by the fundamental theorem of calculus
$$ 
\int_0^T \exp((M-qI)t)\td t = 
(M-qI )^{-1}\left(\exp((M-qI)T)-I
\right)
$$
and, since the real part of the spectrum of the matrix
$M-qI$
is strictly negative by Lemma~\eqref{lem:spectrum}, in the limit as 
$T\to\infty$
we obtain
$$ 
\int_0^\infty \exp((M-qI)t)\td t = 
(qI-M)^{-1}.
$$
\end{pr}

\subsection{(Double) phase-type distributions}
\label{app:Phase_type}
In this section we review basic properties of phase-type distributions, 
as these will play an important role in the sequel.
We refer to Neuts \cite{Neuts} and Asmussen \cite{Asmussenruin} 
for further background on phase-type distributions.

A distribution function
$F:\RR_+\to [0,1]$
is called
\textit{phase-type}
if it is a distribution of the absorption time of
a continuous-time Markov chain on
$(m+1)$
states, for some $m\in\mathbb N$, 
with one state absorbing and the remaining states
transient.
The distribution
$F$
is uniquely determined by the matrix
$A\in\RR^{m\times m}$,
which is the generator of the chain restricted to
the transient states,
and the initial distribution of the chain
on the transient states
$\alpha\in\RR^m$
(i.e. the coordinates
of
$\alpha$
are non-negative and the inequalities
$0\leq\alpha'\1\leq1$
hold, where
$\1$
is the
$m$-dimensional vector with each coordinate equal to one
and
$'$
denotes transposition).
The notation
$X\sim PH(\alpha,A)$
is commonly used for a random variable 
$X$
with cumulative distribution function
$F$. 
Note alse that the law of
the original chain on the entire state space
is given by
$$
\text{the initial distribution}\quad
\begin{pmatrix}
\alpha\\
1-\alpha'\1
\end{pmatrix}
\quad
\text{and the generator matrix}\quad
\begin{pmatrix}
A & \left(-A\right)\1 \\
\mathbf{0} & 0
\end{pmatrix},
$$
where
$\mathbf{0}$
denotes a row of
$m$
zeros.
It is clear from this representation that the cumulative distribution
function
$F$
and its density
$f$
are of the form
\begin{eqnarray}
\label{eq:PTdensity}
F(t)=1-\alpha'e^{tA}\1 \quad\text{and}\quad f(t)=-\alpha'e^{tA}A\1 \quad\text{for any}\quad t\in\RR_+.
\end{eqnarray}
Note also that
$0$
is an atom of the distribution if and only if
$\alpha'\1<1$
in which case
the function
$f$
is a
densitiy of a sub-probability measure on
$(0,\infty)$.
The 
$n$-th moment of the random variable 
$X\sim PH(\alpha,A)$
is given by
$$
\EE\left[X^n\right] =  n!\> \alpha' \left(-A\right)^{-n}\1.
$$

It follows from the definition
the phase-type distribution
that
the matrix
$A$
can be viewed as a generator of a killed continuous-time
Markov chain on
$m$
states.
Therefore we can express the matrix $A$
as
$A=Q-D$,
where
$Q$
is the generator of a chain on
$m$
states and
$D$
is a diagonal matrix with non-negative diagonal elements
that are equal to the coordinates
of the vector
$-A\1$.
Lemma~\ref{lem:spectrum}
therefore implies that the real part of each eigenvalue
of
$A$
is non-positive.
The next proposition 
gives a characterisation of the existence of exponential moments
of a phase-type distribution in terms of the eigenvalues of the matrix
$A$.

\begin{prop}
\label{prop:Phase-typeExpMom}
Let
$X\sim PH(\alpha,A)$
be a phase-type random varaible as defined above
and let
$\lambda_0$
be the eigenvalue of
the matrix
$A$
with the largest real part,
i.e.
$\Re(\lambda_0)=\max\{\Re(\lambda)\>:\> \lambda\>\text{ eigenvalue of }\>A\}.$
Then, for any $u\in\CC$, 
the exponential moment
$\EE[\exp(uX)]$
exists and is finite if and only if
$\Re(u)<-\Re(\lambda_0)$
in which case the following formula holds
$$\EE[\exp(uX)]= \alpha'(A+uI)^{-1}A\1+
(1-\alpha'\1),
$$
where
$I$
denotes an
$m$-dimensional dentity matrix.
\end{prop}

\begin{pr}
It is clear that the identity
\begin{eqnarray}
\label{eq:CharFunPT}
\EE[\exp(uX)]=\PP(X=0)+\int_0^\infty \exp(tu)f(t)dt
\end{eqnarray}
must hold
for all
$u\in\RR$
where
$f$
is the density of
$X$
on the interval
$(0,\infty)$.
Hence the question of existence
of
$\EE[\exp(uX)]$
is equivalent to
the question of convergence of the integral.
Using Formula~\eqref{eq:PTdensity}
for the density
$f$,
the fact
$\exp(t(A+uI))=\exp(tA)\exp(tu)$
for all
$u\in\CC$
and the Jordan canonical decomposition of the matrix
$A$
we
can conclude that
$$\EE[|\exp(uX)|]<\infty\quad\iff \quad-\alpha'\left(\int_0^\infty\exp((A+\Re(u)I)t)dt\right)\> A\1<\infty
\quad\iff\quad \Re(\lambda_0+u)<0,
$$
where
$\lambda_0$
is as defined above. This proves the equivalence in the proposition.

Note that the condition
$\Re(u)<-\Re(\lambda_0)$
implies, by Lemma~\eqref{lem:spectrum}, that the matrix
$A+uI$
is invertible. For any
$T\in\RR_+$
the fundamental theorem of calculus therefore yields
the matrix identity
\begin{eqnarray}
\label{eq:IntRepForMatrices}
\int_0^T\exp((A+uI)t) dt\>\>=\>\>(A+uI)^{-1}\left[\exp((A+uI)T)-I\right].
\end{eqnarray}
Since all the eigenvalues of
$A+uI$
have a strictly negative real part, it follows from Jordan canonical
decomposition of
$A+uI$
that
$\lim_{T\to\infty}\exp((A+uI)T)=0$.
Therefore identities~\eqref{eq:CharFunPT}
and~\eqref{eq:IntRepForMatrices}
conclude the proof of the proposition.
\end{pr}

More generally, a {\it double phase-type} jump distribution 
$DPH(p,\beta^+, B^+,\beta^-,B^-)$ is defined to have density
\begin{eqnarray}
\label{eq:Jump_Density}
 f(x) &  :=  & pf^+(x)
I_{(0,\infty)}(x) +
(1-p)f^-(-x)
I_{(-\infty,0)}(x) \quad\text{such that}\\ 
& & p\in[0,1],\quad
f^\pm\sim PH(\beta^\pm,B^\pm),\quad
f^\pm(x)=-(\beta^\pm)'e^{xB^\pm}B^\pm\1 
\quad\text{and}\quad
\1'\beta^{\pm}=1,\nonumber
\end{eqnarray}
where
the phase-type distributions 
$PH(\beta^\pm,B^\pm)$
are
as described above, 
$\1$
is a vector of the appropriate size with all coordinates equal to
$1$
and
as usual 
$I_A$
denotes
the indicator of the set
$A$.
The condition
$\1'\beta^{\pm}=1$
ensures that the distribution of jump sizes has no atom at zero.

The class of 
double phase-type distributions is vast.  Not only does it contain 
double exponential distributions
\begin{eqnarray}
\label{eq:DExp_Density}
f(x) & := & p\alpha^+e^{-x\alpha^+}I_{(0,\infty)}(x) +
(1-p)\alpha^-e^{x\alpha^-}I_{(-\infty,0)}(x)\quad\text{where}
\quad \alpha^\pm>0\text{ and } p\in[0,1],
\end{eqnarray}
mixtures of double exponential distributions and Erlang distributions
but this class is in fact dense in the sense of weak convergence
in the space of all probability distributions on
$\RR$. 

\begin{prop}\label{prop:dense} 
Let $F$ be a probability distribution function on $\mathbb R$. 
Then there exists a sequence $(F_n)_{n\in\mathbb N}$ of double-phase-type distributions $F_n$ such that $F_n\Rightarrow F$ as $n\to\infty$.\footnote{We write $F_n\Rightarrow F$ for a sequence of distribution functions $F_n$ and a distribution function $F$ if $F_n$ converges in distribution to $F$, that is, $\lim_{n\to\infty}F_n(x)= F(x)$ for all $x$ where $F$ is continuous.}
\end{prop}
This result directly follows from the three observations that (a) any probability distribution on the real line can be approximated in distribution arbitrarily closely by a random variable taking only finitely many values and (b) 
any constant random variable is the limit in distribution of Erlang or the negative of Erlang random variables, and (c) a mixture of Erlang distributions is a phase-type distribution.

An important property of exponential distributions 
is the lack-of-memory property, which can be 
generalised to stopping times as follows:

\begin{lem}
\label{lem:ExpGen}
Let
$(\FF_t)_{t\geq0}$
a filtration and let
$\rho$
be any stopping time\footnote{
By definition the stopping time 
$\rho$
takes values in 
$[0,\infty]$
and satisfies the condition
$\{\rho\leq t\}\in\FF_t$
for all
$t\in[0,\infty)$.
The 
$\sigma$-algebra
$\FF_\rho$
consists of all events
$A$
such that 
$A\cap\{\rho\leq t\}\in\FF_t$
for all
$t\in[0,\infty)$.}
with respect to 
this filtration. Let
$\mathbf e_q$
be an exponentially distributed random variable 
with parameter 
$q>0$
which is independent of the 
$\sigma$-algebra
generated by
$\cup_{t\geq0}\FF_t$.
Then 
the equality 
$$ 
\EE\left[I_{\{\rho<\mathbf e_q\}}\exp(-\lambda(\mathbf e_q-\rho))\Big\lvert\FF_\rho\right]=
\frac{q}{\lambda+q} \te{-q\rho} \quad
\text{holds for all}\quad\lambda\geq0
$$
and hence
the positive random variable
$\mathbf e_q-\rho$
defined on the event 
$\{\rho<\mathbf e_q\}$
is,
conditional on 
$\FF_\rho$,
exponentially distributed
with parameter
$q$.
\end{lem}
{\bf Remarks.} (i) This lemma can be viewed as a generalisation of the lack of memory 
property, 
$$\PP(\mathbf e_q>t+s\lvert \mathbf e_q>s)=\PP(\mathbf e_q>t),$$
of the exponential random variable 
$\mathbf e_q$
when the constant time
$s$
is substituted by a stopping time
$\rho$.
Note also that it follows from the lemma 
that the conditional probability of 
the event
$\{\rho<\mathbf e_q\}$
equals
$$\PP(\mathbf e_q>\rho\lvert\FF_\rho)=\exp(-q\rho).$$

(ii) Phase-type distributions enjoy a similar property 
that can be seen as a generalisation of 
the lack-of-memory of the exponential distribution.
More specifically, let $T$ follow a $\mrm{PH}(\alpha, B)$ 
distribution independent of the $\sigma$-algebra generated by 
$\cup_{t\geq0}\FF_t$. 
Than for any stopping time $\rho$ with respect to 
$(\FF_t)_{t\ge 0}$,
the random variable
$T-\rho$ 
defined on the event 
$\{\rho<T\}$, conditional on $\mc F_\rho$, is 
$\mrm{PH}(\alpha_\rho, B)$ distributed where 
$$\alpha_\rho = (\alpha' \te{\rho B}\1)^{-1}\alpha'\exp\{\rho B\},$$
since the identity 
$$\EE\left[I_{\{\rho<T\}}\exp(-\lambda(T-\rho))\Big\lvert\FF_\rho\right]=
\alpha' \exp\left(\rho B\right)
\left(B-\lambda I\right)^{-1} B\1
$$
holds for all
$\lambda\geq0$.
This follows by the same argument as in the proof of Lemma~\ref{lem:ExpGen}. 
Furthermore we have the following expression for the conditional probability
of the event
$\{\rho<T\}$:
$$\PP(T>\rho\lvert\FF_\rho)=
\alpha' \exp\left(\rho B\right)\1.
$$

\begin{pr}
The following direct calculation based on Fubini's theorem,
which is applicable since
all the functions are non-negative,
$$\EE\left[I_AI_{\{\rho<\mathbf e_q\}}\exp(-\lambda(\mathbf e_q-\rho)) \right]=
\EE\left[I_AI_{\{\rho<\infty\}}e^{\lambda\rho} q \int_\rho^\infty e^{-(\lambda+q) t}dt \right]
= \EE\left[I_A\frac{q}{\lambda+q}e^{-q\rho}\right],\quad\text{where}\quad
A\in\FF_\rho,
$$
proves the identity in the lemma
for all non-negative
$\lambda$.
Since the Laplace transform uniquely determines the distribution
of a random variable
the lemma follows.
\end{pr}

\section{Stochastic volatility models with jumps}
\label{sec:Model}
We next describe in detail the two classes of stochastic volatility 
models with jumps that we will consider. 

Let $v=\{v_t\}_{t\ge 0}$ be a Markov process that takes positive values,
modelling the underlying stochastic variance, and let $X$ be a L\'{e}vy process\footnote{A L\'{e}vy process $X=\{X_t\}_{t\ge 0}$ is a stochastic process that has independent and stationary increments, and has right-continuous paths with left-limits with $X_0=0$.}
which drives the noise in the log-price process.
The processes are taken to be 
mutually independent and are both defined on some probability 
space $(\Omega,\mc F,\PP)$. 

The law of $X$ is determined by its 
characteristic exponent $\psi:\mathbb R\to\mathbb R$ 
which is according to the L\'{e}vy-Khintchine formula given by
\begin{equation}\label{eq:lk}
\EE[\te{\cu u X_t}]=\te{t\psi(u)} 
\end{equation}
with
\begin{equation}\label{eq:lk2}
\psi(u) = \cu cu - \frac{\sigma^2}{2}u^2 + \int_{-\infty}^\infty[\te{\cu u x} - 1 - \cu u x I_{\{|x|\leq 1\}}]\nu(\td x),
\end{equation}
where $\sigma^2\ge 0$ and $c$ are constants and $\nu$ is the L\'{e}vy measure
that satisfies the integrability condition 
$\int_{\mathbb R} (1\wedge x^2)\nu(\td x) < \infty.$ The triplet 
$(c,\sigma^2,\nu)$ is also called the characteristic triplet of $X$.

To guarantee that the option prices be finite we impose 
the usual restriction that $X$ admits (positive) exponential moments; 
more precisely, we assume that for some $p> 1$
\begin{equation}\label{exp:cond}
\int_1^\infty \te{px}\nu(\td x) < \infty,
\end{equation}
which implies that $\EE[\te{p X_t}]<\infty$ for all $t\ge 0$.
In this case the identity \eqref{eq:lk} remains valid for all $u$ 
in the strip 
$\{u\in\mathbb C: \Im(u) \in (-p,0]\}$
in the complex plan, where the function $\psi$ is analytically 
extended to this strip.

In this setting a candidate stochastic volatility process with jumps 
$S=\{S_t\}_{t\in[0,T]}$, where $T>0$ denotes a maturity 
or time-horizon, is given by
\begin{eqnarray}
\label{eq:Model2}
S_t:=S_0\exp\left((r-d)t + \int_0^t\sqrt{v_u} \td X_u - \int_0^t \psi(-\cu \sqrt{v_s})\td s\right),\qquad S_0=s>0,
\end{eqnarray}
where 
$r$
and
$d$
are the instantaneous interest rate and dividend yield respectively.
Here we assume that the variance process $v$ satisfies the 
following integrability condition:
\begin{equation}
\label{eq:IntegrabCond}
\int_0^T |\psi(-\cu \sqrt{v_s})|\td s < \infty\quad a.s.
\end{equation}
It is easy to see
by conditioning on the filtration generated by
the variance process
$v$
that the integrability condition
in~\eqref{eq:IntegrabCond}
implies the martingale property of the discounted process
$\{\te{-(r-d)t}S_t\}_{t\in[0,T]}$.

Note that if we take for example $X$
to be a Brownian motion with drift and
$v$ an independent square-root process, the process
$S$ reduces to a Heston model with zero correlation between the driving
Brownian factors (see e.g.~\cite{Gatheral}). 
The class of models described by~\eqref{eq:Model2}
is quite flexible, and contains for example the 
stochastic volatility models with jumps
described in Lipton~\cite{Lipton_Book},
as long as there is no correlation between the driving Brownian motions.

A related class of models that has 
been proposed in the literature is the one 
where the effect of stochasticity of volatility is achieved by randomly changing the time-scale 
(see e.g. Carr et al.~\cite{CarrMadanGemanYor_StochVolLevy}); 
in the setting above the price process $\{S_t\}_{t\in[0,T]}$
is defined by
\begin{eqnarray}
\label{eq:Model_Gang2}
S_t:=S_0\exp\left((r-d)t + X_{V_t} - \psi(-\cu) V_t\right)
,\qquad S_0=s>0,
\qquad\text{where}\quad V_t:=\int_0^tv_u \td u,
\end{eqnarray}
and we assume that $v$ satisfies the integrability condition
\begin{equation}
V_T < \infty \quad a.s.
\end{equation}
Also in this case the discounted process $\{\te{-(r-d)t}S_t\}_{t\in[0,T]}$ is a martingale.

It is clear from the definitions that in the case
where $X$
is a Brownian motion with drift, the classes of models in~\eqref{eq:Model2}
and~\eqref{eq:Model_Gang2}
coincide, due to the scaling property of Brownian motion. 
Whereas 
the effect of the variance process $v$ on the Brownian motion with drift is the same in
both classes of models, the effect of the process $v$ on the behaviour of
jumps is different. In~\eqref{eq:Model_Gang2} 
the Markov process $v$ modulates only the intensity of the jumps of
$X$ while in model~\eqref{eq:Model2}
the volatility scales the distribution of size of the jumps
but does not affect the intensity.

In the next section we will describe a modelling framework in which
any model in the classes given by~\eqref{eq:Model2}
and~\eqref{eq:Model_Gang2} can be approximated. 
The approximation 
in Section \ref{sec:appr} 
retains the structural properties required for the semi-analytic
pricing (i.e. up to an integral transform) of barrier
options, forward starting options and volatility derivatives.

\subsection{A class of regime-switching models}

Let the set
$E^0:=\{1,\ldots,N_0\}$
be the state-space of a continuous-time Markov
chain
$Z=(Z_t)_{t\geq0}$
and let the
the process
$W=(W_t)_{t\geq0}$
denote
a standard Brownian motion which is independent
of the chain
$Z$.
For each
$i\in E^0$
let the process
$J^i:=(J^i_t)_{t\geq0}$
be a compound Poisson process with intensity
$\lambda_i\geq0$ and jump-sizes distributed 
according to a double-phase-type distribution 
$DPH(p_i,\beta_i^+, B_i^+,\beta_i^-, B_i^-)$.
In particular the jump-size distributions have no
atom at zero, i.e. 
$(\beta_i^+)'\1=1$
for all
$i\in E^0$
such that 
$\lambda_ip_i>0$
and analogously for
$\beta_i^-$.
Assume further that the processes
$J^i$
are mutually independent as well as independent from the Brownian
motion
$W$
and the chain
$Z$.

In this setting consider the following model
for the underlying price process
$S=(S_t)_{t\ge0}$,
the (domestic) money market account
$B^D=(B_t^D)_{t\ge0}$
and the cumulative dividend yield
$B^F=(B_t^F)_{t\ge0}$:
\begin{eqnarray}
\label{eq:DomesticBond}
B_t^D & := & \exp\left(\int_0^t R_D(Z_s)\td s\right), \quad
B_t^F  :=  \exp\left(\int_0^t R_F(Z_s)\td s\right),\quad
S_t :=  \exp(X_t),
\end{eqnarray}
where 
\begin{eqnarray}
X_t & := & x+\int_0^t\mu(Z_s)\td s+\int_0^t\sigma(Z_s)\td W_s+\sum_{i\in E^0}\int_0^t I_{\{Z_s=i\}}\td J_s^i.
\label{eq:Process}
\end{eqnarray}
In the case of the Foreign Exchange market the process
$B^F$
can be interpreted as a foreign money market account.
The point
$x\in\RR$
is the starting value of the process
$X$
and $R_D,R_F, \mu,\sigma:E^0\to \RR $
are given real-valued functions
on $E^0$ such that $R_D,R_F$ are non-negative
and 
$\sigma$
is strictly postitive.
%
%
To price derivatives in our model, we need to understand the 
law of the Markov process
$(X,Z)$,
which 
is determined by the characteristic matrix exponent 
$K$, defined as follows.

\begin{defin}
The {\em characteristic matrix exponent} 
$K: \mathbb R\to \mathbb C^{N_0\times N_0}$ 
of $(X,Z)$ is 
given by
$$
K(u) = Q + \Lambda(u),
$$
where $Q$
denotes the generator of the chain
$Z$ and, for $u\in\mathbb R$, 
$\Lambda(u)$ is a diagonal matrix
of size
$N_0\times N_0$,
where the 
$i$-th diagonal element equals
the characteristic exponent
of
the process
$X$
in regime
$i$,
given by
\begin{equation}\label{eq:psii}
\psi_i(u):=\cu u\mu_i - \sigma_i^2u^2/2+
\lambda_i\left[p_i(\beta_i^+)'(B_i^++\cu uI)^{-1}B_i^+\1+ (1- p_i)(\beta_i^-)'(B_i^--\cu uI)^{-1}B_i^-\1-1\right].
\end{equation}
where 
$I$ and $\1$
are an identity matrix and
a vector with all coordinates 
equal to one
of the appropriate dimensions.
\end{defin}

\noindent \textbf{Remarks.} (i) Note that the functions $\psi_i$ 
defined in~\eqref{eq:psii} can be analytically 
extended to the strip in the complex plane $\Im(u)\in(-\alpha_i^+,\alpha_i^-)$
where
\begin{eqnarray}
\label{eq:SmallestEigVal}
\alpha_i^\pm:=\min\{-\Re(\lambda)\>:\> \lambda\>\text{ eigenvalue of }\>B_i^\pm\}
\quad\text{ for any state }\quad i\in E^0.
\end{eqnarray}

(ii) In the special case that the jumps follow a 
double exponential distribution
the diagonal elements of the matrix
$\Lambda(u)$
take the simpler form 
$$\psi_i(u)=\cu u\mu_i - \sigma_i^2u^2/2+
\lambda_ip_i\left(\frac{\beta_i^+}{\beta_i^+-\cu u}-1\right)+
\lambda_i(1- p_i)\left(\frac{\beta_i^-}{\beta_i^-+\cu u}-1\right),$$
where 
$\beta_i^\pm$ and $p_i$ are the parameters of the 
double exponential distribution.

(iii) Throughout the paper we will use
$\EE_{x,i}[\cdot]$
to denote the conditional expectation 
$\EE[\cdot|X_0=x,Z_0=i]$
and on occasion 
$\EE_{i}[\cdot]$
to represent
$\EE_{0,i}[\cdot]$.

We now define two matrices the will play an important role in the sequel.

\begin{defin}
The {\em discount rate matrix} $\Lambda_D$ is
the diagonal matrix with elements $\Lambda_D(i,i):= R_D(i)$
where
$i\in E^0$.
The \textit{dividend yield matrix}
$\Lambda_F$ is the diagonal matrix given by 
$\Lambda_F(i,i):= R_F(i)$
for
$i\in E^0$.
\end{defin}

\begin{thm}
\label{thm:CharFun}
The discounted characteristic function of the Markov process
$(X,Z)$
is given by  the formula
\begin{equation}\label{eq:expux}
\EE_{x,i}\left[\frac{\exp(\cu uX_t)}{B_t^D}I_{\{Z_t=j\}}\right]=
\exp(\cu u x)\cdot \exp(t(K(u)-\Lambda_D))(i,j)
\end{equation}
for all $u\in\mathbb R$.
\end{thm}
{\bf Remarks.} (i) The left-hand side is finite for all $u\in\mathbb C$ in the strip 
$\Im(u)\in(-\alpha_*^+,\alpha_*^-)$ where 
\begin{equation}
\alpha^+_*=\min\{\alpha^+_k:\lambda_k p_k > 0, k\in E^0\}\q\text{and}\q 
\alpha^-_*=\min\{\alpha^+_k:\lambda_k (1-p_k) > 0, k\in E^0\},
\end{equation}
the quantities $\alpha_i^\pm$ are defined in \eqref{eq:SmallestEigVal} and the minimum over the empty set is taken to be $+\infty$.
It follows, by analytical continuation, 
that the identity \eqref{eq:expux} remains valid 
for all $u$ in this strip.  Furthermore, the 
slope of the implied volatility smile in model~\eqref{eq:Process}
is determined by $\alpha^+_*$ and $\alpha^-_*$ 
(see Subsection~\ref{subsec:IV}).

(ii) The Markov property and Theorem~\ref{thm:CharFun} imply that the process
$\{S_tB^F_t/B^D_t\}_{t\geq0}$
is a martingale if the following two conditions hold:
\begin{eqnarray}
\label{mc1}
1 &<&  \alpha^+_k,\quad \text{for all $k\in E^0$ such that $\lambda p_k>0$},\\
\Lambda(-\cu) &=& \Lambda_D-\Lambda_F.
\label{mc2}
\end{eqnarray}
Condition \eqref{mc1} ensures that $\EE_{i,x}[S_T]$
is finite for
all
$T\geq0$
and hence by Theorem~\ref{thm:CharFun} takes the form
$\EE_{i,x}[S_T]=\te{x}\left[\exp\left(T K(-\cu)\right)\1\right](i)$. 
The equality in~\eqref{mc2} guarantees that $S$ has instantaneous drift
given by the rates $\Lambda_D-\Lambda_F$. 
Any model from the class \eqref{eq:DomesticBond}--\eqref{eq:Process} 
that satisfies conditions 
\eqref{mc1} and \eqref{mc2} can be taken as a specification of the 
price process of the risky-asset under a pricing measure. 
From now on we assume that model~\eqref{eq:Process} is specified
under the pricing measure given by condition~\eqref{mc1}--\eqref{mc2}.

(iii) For later reference we record that, under a pricing measure, the price at time $s$ of a zero coupon bond maturing at time $t\ge s$ is given by
\begin{equation}
\EE_i\le[\frac{1}{B_t^D}\bigg|\mc F_s^{(X,Z)}\ri] = \frac{1}{B_s^D}
\cdot\le(\exp((t-s)(Q-\Lambda_D))\mathbf 1\ri)(Z_s),
\end{equation}
where $\mc F_s^{(X,Z)}=\s\{(X_u,Z_u)\}_{u\leq s}$ denotes the standard 
filtration generated by $(X,Z)$.
In particular, at time $0$ the price is given by 
$$\EE_i\le[\le(B_t^D\ri)^{-1}\ri]=\le(\exp(t(Q-\Lambda_D))\mathbf 1\ri)(i).$$

(iv) The infinitesimal generator $\mathcal L$ 
of the Markov process $(X,Z)$ acts on sufficiently smooth functions\footnote{For example, functions $f$ with $f(\cdot,i)\in C^2_c(\mathbb R)$ for $i\in E^0$, where
$C^2_c(\mathbb R)$ are the twice continuously differentiable functions with compact support.} $f: \mathbb R\times E^0\to \mathbb R$ as 
\begin{eqnarray}\nn
\mathcal Lf(x,i) &=& \frac{\sigma^2(i)}{2}f''(x,i) + \mu(i) f'(x,i) + 
\lambda(i)  \left[ \int_{\mathbb R} f(x+z,i) g_i(z)\td z-f(x,i)\right] \\ &+& \sum_{j\in E^0} q_{ij}[f(x,j) - f(x,i)],
\label{eq:generator}
\end{eqnarray}
where $g_i$ is the density of the double phase-type distribution $DPH(p_i,\beta_i^+, B_i^+,\beta_i^-, B_i^-)$, $q_{ij}$ is the $ij$th element of $Q$ and $'$ denotes differentiation with respect to $x$.

(v) For a specific regime-switching model 
(namely the case where the Markov chain $Z$ 
has two states only) the calibration is studied in~\cite{Pistorius:2009}.

\begin{pr}
It is clear from the definition of 
$(X,Z)$
that it is a Markov process. Let 
$\FF_t^Z:=\sigma(Z_s:s\in[0,t])$
be the 
$\sigma$-algebra generated by the chain
$Z$
up to time
$t$.
Since the compound Poisson processes 
and Brownian motion in model~\eqref{eq:Process} are mutually 
independent as well as independent of the 
$\sigma$-algebra
$\FF_t^Z$,
it is easy to see that by conditioning on 
$\FF_t^Z$ 
for any
$i\in E^0$
we obtain
\begin{eqnarray}
\EE_{x,i}\left[\exp(uX_t)|\FF_t^Z\right] & = & \exp\left(ux+u\int_0^t\mu(Z_s)\td
s+\frac{u^2}{2}\int_0^t\sigma(Z_s)^2\td s+\int_0^t \nu(Z_s,u)\td s\right),
\label{eq:CondExpectMain}
\\
\nu(i,u) & := &\lambda_i \left[\EE[\exp(uJ_i)]-1\right]\nonumber \\
& = & \lambda_i\left[p_i(\beta_i^+)'(B_i^++uI)^{-1}B_i^+\1+ (1- p_i)(\beta_i^-)'(B_i^--uI)^{-1}B_i^-\1-1\right],\quad
\nonumber
\end{eqnarray}
where the random variable 
$J_i$
denotes the size of jumps 
of the compound Poisson process
$J^i$.
The last equality in this calculation is a consequence of the choice~\eqref{eq:Jump_Density}
of the distribution of
jump sizes
and Proposition~\ref{prop:Phase-typeExpMom}.
Therefore the complex number
$u$
must be contained in all intervals 
$(-\alpha_k^-,\alpha_k^+)$,
$k\in E^0$,
where 
$\alpha_k^\pm$
are defined in Theorem~\ref{thm:CharFun}.
The identity in~\eqref{eq:CondExpectMain}
holds more generally
for any jump-distribution that 
admits a moment generating function.
The well-known identity 
from the theory of Markov chains 
given in Lemma~\ref{lem:MC_Semigroup}
can now be applied to 
obtain the expectations of the 
expressions on both sides of~\eqref{eq:CondExpectMain}.
This concludes the proof of Theorem~\ref{thm:CharFun}.
\end{pr}



From Theorem~\ref{thm:CharFun} one may obtain an explicit expression for 
the marginal distributions of $(X,Z)$ by inverting the Fourier transform \eqref{eq:expux}:

\begin{prop}
\label{prop:ExisteceOfDensity}
For any $T>0$, the joint distribution 
$q^{x,i}_T(y,j) = \frac{\td }{\td y} \PP_{x,i}[X_T\leq y, Z_T = j]$
 of $(X_T,Z_T)$ is given by
\begin{eqnarray}
\label{eq:ForwardSmileVector}
q^{x,i}_T(y,j) & = & \frac{1}{2\pi}\int_\RR e^{\cu
\xi(x-y)}\exp\left(K\left(\xi\right)T\right)(i,j)\>\td\xi,
\quad y\in\RR, i,j\in E^0,
\end{eqnarray}
In particular, $X_T$ is a continuous random variable 
with probability density function $q^{x,i}_T(y) = \frac{\PP_{x,i}[X_T\in \td y]}{\td y}$
given by
\begin{eqnarray}
\label{eq:DensityOfX}
q^{x,i}_T(y)
& = & 
\frac{1}{2\pi}\int_\RR e^{\cu \xi(x-y)} \le[\exp\left(K\left(\xi\right)T\right)\1\ri](i)\>\td\xi,\quad y\in\RR, i\in E^0.
\end{eqnarray}
\end{prop}

\begin{pr}
It is well-known that a probability law on the real line 
$\RR$
has a density with respect to the Lebesgue measure if its characteristic function 
is in
$L^1(\RR)$.
The characteristic function of 
$X_T$
by Theorem~\ref{thm:CharFun}
equals
\begin{equation}\label{eq:ch}
\EE_{x,j}\left[\exp(\cu\xi X_T)\right]=e^{\cu\xi x} [\exp(K(\xi)T)\1](j).
\end{equation}
We now show that this characteristic function is asymptotically equal to
$\exp(-c \xi^2)$,
as
$|\xi|\to\infty$,
for some positive constant 
$c$.
Note first that the volatility vector
$\sigma$
in model~\eqref{eq:Process}
has non-zero coordinates by assumption and the spectra of matrices
$B_i^\pm$,
$i=1,\ldots,N_0$,
do not contain any points  
of the form
$\cu\xi$,
for 
$\xi\in\RR$,
by Lemma~\ref{lem:spectrum}.
Therefore the functions
$\xi\mapsto\psi_i(\xi)$,
$i=1,\ldots,N_0$,
defined in~\eqref{eq:psii} 
are asymptotically equal to downward facing
parabolas. A further application of 
Lemma~\ref{lem:spectrum}
implies that the characteristic function has the desired 
asymptotic behaviour.
This further implies that the density 
$q^{x,i}_T$
exists and is given by the inversion formula~\eqref{eq:DensityOfX}.

Theorem~\ref{thm:CharFun} and a
 similar argument to the one outlined in the previous paragraph
imply that the function 
\begin{eqnarray*}
\xi & \mapsto & 
e^{\cu\xi x}\exp(TK(\xi))(i,j) =
\EE_{x,i}\left[\exp(\cu\xi X_T)I_{\{Z_T=j\}}\right] 
\end{eqnarray*}
is in 
$L^1(\RR)$
and that the two equalities hold.
Therefore the Fourier inversion formula is valid and
the identity in~\eqref{eq:ForwardSmileVector}
follows.
\end{pr}

\subsection{Two step approximation procedure}\label{sec:appr}
The construction of a regime-switching L\'{e}vy process with  jump sizes distributed according to a double phase-type distribution that approximates a given stochastic volatility process with jumps from either of 
the two-classes \eqref{eq:Model2} and \eqref{eq:Model_Gang2}
takes place in two steps: 
\begin{itemize}
\item[(i)] Approximation of the variance process $v$ by a finite-state
continuous-time Markov chain and 
\item[(ii)] Approximation of the L\'{e}vy process $X$ by a L\'{e}vy process with double-phase-type jumps.
\end{itemize}
By approximating the variance process by a finite state Markov chain 
the resulting approximating process is a regime-switching 
L\'{e}vy process. The approximation of the jump part of $X$ by a compound Poisson process with double phase-type jumps will enable us to 
employ matrix Wiener-Hopf factorisation results, needed to obtain tractable
formulae for the prices of barrier-type options. The two steps will be described in
detail in the
present section.

\subsubsection{Markov chain approximation of the variance process}
The first  step of the approximation procedure that was outlined above is to
approximate the variance process
$v$ by a finite-state continuous-time Markov chain on some grid contained in the
positive real line. 
We will restrict ourselves to the case that the variance process $v$ 
is a Feller process on the state space 
$\mathbb R_+ 
= [0,\infty)$. This assumption implies that $v$ is a Markov 
process satisfying some regularity properties.

The Feller property is phrased in terms of the semi-group 
$(P_t)_{t\ge 0}$ of $v$ acting 
on $C_0(\mathbb R_+)$, the set of continuous 
functions on $\mathbb R_+$ that tend 
to zero at infinity. 
Recall that, for any Borel function $f$ on $\mathbb R_+$ and $t\ge 0$, 
the map $P_t f:\mathbb R_+\to\mathbb R$ is given by
$$P_tf(x) = \EE_x[f(v_t)].$$ 
\begin{As} The Markov process $v=\{v_t\}_{t\ge 0}$ is a Feller process;
that is, for any $f\in C_0(\mathbb R_+)$, the family $(P_tf)_{t\ge 0}$
satisfies the following two properties:
\begin{itemize}
\item[(i)] $P_tf \in C_0(\mathbb R_+)$ for any $t>0$;
\item[(ii)] $\lim_{t\downarrow 0}P_tf(v) = f(v)$ for any $v\in \mathbb R_+$.%
\end{itemize}
\end{As}

An approximating Markov chain $Z$ with generator $Q$ on a 
state-space $E^{0}=\{x_1, \ldots, x_{N_0}\}$ can be constructed 
by choosing $E^{0}$ 
to be some appropriate (non-uniform) grid in $\mathbb R_+$, and 
specifying the generator $Q$ such that an appropriate set of  
instantaneous (local) moments of the chain $Z$ and the 
target process $v$ are matched. 
See~\cite{Pistorius_MarkovBarrier:2009} for details on this procedure.

Denote by $\mc G: \mc D \to C_0(\mathbb R_+)$ 
the infinitesimal generator of $v$ defined on its domain $\mc D$, and 
let $Z^{(n)}$ be a sequence of Markov chains with generators $Q^{(n)}$
and state-spaces $E^{0(n)}=\{x^{(n)}_1, \ldots, x^{(n)}_{N^{(n)}}\}$, and denote by $Q^{(n)} f_n$ the vector with coordinates 
$$Q^{(n)} f_n(x_i) = \sum_{x_j\in E^{0(n)}}Q^{(n)}(x_i,x_j)f(x_j), \qquad x_i\in E^{0(n)}.$$
The sequence $Z^{(n)}$ weakly approximates the variance process $v$ if the range of the  state-spaces $E^{0(n)}$ grows sufficiently fast as $n$ tends to infinity, and if,
for all regular functions $f$,
$Q^{(n)} f_n$ converges uniformly to $\mc Gf$, that is $\epsilon_n(f)\to 0$
where
\begin{equation*}
\epsilon_n(f) := \max_{x\in(E^{0(n)})^o}\le|Q^{(n)} f_n(x) - \mathcal G 
f(x)\ri| 
\end{equation*}
and $(E^{0(n)})^o$ is equal to $E^{0(n)}$
without the smallest and the largest elements.
The precise statement reads as follows:

\begin{thm}\label{thm:conv} 
Assume that the following two conditions are satisfied
for any function in a core\footnote{A core $\mathcal{C}$ 
of the operator
$\mc L$
is a subspace of the domain of 
$\mc L$
that 
is (i) dense in
$C_0(\mathbb R_+)$ 
and (ii) there exists
$\lambda>0$
such that the set 
$\{(\lambda - \mathcal L)f: f\in \mc C\}$ is dense in $C_0(\mathbb R_+)$.}
of 
$\mathcal L$:
\begin{eqnarray}
& &
\label{eq:ConvHom}
\e_n(f)\to 0 \q\text{as $n\to\infty$},
\\
& & \text{either (i) $\lim_{y\searrow0}\mc Gf(y)=0$ or 
(ii) $\lim_{n\to\infty}\PP_x\left[\tau_{(E^{0(n)})^o}^{(n)}>T\right]=1$},
\label{eq:CorrectCondition}
\end{eqnarray}
where for any set
$G\subset\mathbb R_+$
we define
$\tau_G^{(n)}=\inf\{t\ge0: X^{(n)}_t\notin G\}$.
Then it holds that, as $n\to\infty$, 
$v^{(n)}
\stackrel{\mathcal L}{\Rightarrow} v$.\footnote{By $v^{(n)}\stackrel{\mc L}{\Rightarrow}v$ we denote the convergence in law of $v^{(n)}$ to $v$ in the Skorokhod topology, i.e. convergence of the distributions of $v^{(n)}$ to those of $v$ in the set of probability measures on the Skorokhod space endowed with the Skorokhod topology.}
%
\end{thm}
{\bf Remark.} (i) Note that the convergence in law implies in particular that 
\begin{eqnarray*}
\EE_x\left[g\le(Z^{(n)}_{T}\ri)\right]&\longrightarrow&
\EE_x\le[
g(v_{T})\ri]
\end{eqnarray*}
for any bounded continuous function $g$.

(ii) A proof of this statement can be found in~\cite{Pistorius_MarkovBarrier:2009}.



\subsubsection{Approximation of a L\'{e}vy process} 
The second stage of the aforementioned approximation procedure 
amounts to an approximation of a L\'evy process by a
compound Poisson process with double phase-type jumps. 

\begin{prop}
For any L\'{e}vy process $X$ there exists a sequence $(X^{(n)})_{n\in\NN}$
of L\'{e}vy processes with double phase-type jumps such that 
$X^{(n)}\stackrel{\mc L}{\Rightarrow} X$ as $n\to\infty$.%
\end{prop}
{\bf Remarks.} (i) A proof of this result can be found in e.g. 
Jacod and Shiryaev \cite[Section VII.3]{JacodShiryaev}. 
It is based on the fact that a sequence $(X^{(n)})_{n\in\NN}$ of L\'{e}vy processes weakly converges to a given L\'{e}vy process $X$ if and only if $X^{(n)}_1$ converges 
in distribution to $X_1$ (see e.g. \cite[Corollary VII.3.6]{JacodShiryaev} for a proof). 

(ii) Furthermore \cite[Corollary VII.3.6]{JacodShiryaev} implies that
a sufficient condition to guarantee that  $X^{(n)}_1$ converges in distribution to
$X_1$ is that the characteristic triplets $(c_n,\sigma^2_n,\nu_n)$ of $X^{(n)}$
converges to the triplet $(c,\sigma^2,\nu)$ of $X$ as follows as $n\to\infty:$ for some
$a>0$ that is a continuity point of $\nu(\td x)$ and $\nu(-\td x)$ it holds that
\begin{eqnarray}
\label{cnsn} c_n \to c, \qquad \sigma^2_n + \int_{(-a,a)} x^2\nu_n(\td x)  &\to& 
\sigma^2 + \int_{(-a,a)}x^2\nu(\td x) \\
\int_{(0,\infty)}(x^2\wedge a)|\ovl\nu_n(x)-\ovl\nu(x)|\td x
&+& \int_{(-\infty,0)}(|x|^2\wedge a)|\unl\nu_n(x)-\unl\nu(x)|\td x
\to 0 \label{eq:nunnu}
\end{eqnarray}
where, for any measure $m$ on $\mathbb R$, $\ovl m$ and $\unl m$ are the left and right tails, $\ovl m(x)=m([x,\infty))$, $\unl m(x) = m((-\infty,x])$.

Suppose now that the L\'{e}vy process $X$ is a model for the log of a stock price. Then
$X$ has a triplet $(c,\sigma^2,\nu)$ satisfying the exponential moment condition in~\eqref{exp:cond}. 
An example of a sequence of L\'{e}vy processes with DPH distributed jumps that weakly converges to $X$ is then 
given as follows.
Let $\lambda_n=\nu((-1/n,1/n)^c)$ and $F_n$ be a double phase-type distribution that approximates in distribution the probability measure $\WT F_n(\td x) = I_{\{|x|\ge1/n\}}\nu(\td x)/\lambda_n$, and define the measure $\nu_n$ by $\nu_n(\td x) = \lambda_n F_n(\td x)$. Here the $F_n$ and $\sigma^2_n$ are to be chosen such that 
\eqref{eq:nunnu}
and the second requirement in \eqref{cnsn} hold true.
See e.g.~\cite{PhaseType_Fit} for a fitting procedure based on the EM algorithm.
Then the sequence of L\'{e}vy processes $X^{(n)}$ with triplets 
$(c,\sigma^2_n, \lambda_n F_n)$ satisfies the conditions \eqref{eq:nunnu} 
and thus approximates $X$ in law as $n$ tends to infinity.

\subsection{Convergence of the approximation procedure}

Combining the two steps in the approximation we can now identify 
candidate sequences of regime-switching processes that converge to 
either of the stochastic volatility processes 
with jumps \eqref{eq:Model2} and \eqref{eq:Model_Gang2}
and establish the convergence. 

Let $(X^{(n)})_{n\in\NN}$ be a sequence of L\'{e}vy processes with DPH jumps, and let
$(Z^{(n)})_{n\in\NN}$ 
be a sequence of Markov chains 
that is independent of $X$ and $(X^{(n)})_{n\in\NN}$.
Let $\psi_n$ denote the characteristic exponent of $X^{(n)}$, and $(c_n,\sigma^2_n,\nu_n)$ the characteristic triplet.
Consider then the sequences of stochastic 
processes $(S^{(int-n)})_{n\in\NN}$ and 
$(S^{(tc-n)})_{n\in\NN}$ 
with 
$S^{(int-n)}=\{S^{(int-n)}_t\}_{t\in[0,T]}$ 
and 
$S^{(tc-n)}=\{S^{(tc-n)}_t\}_{t\in[0,T]}$
given by
\begin{eqnarray*}
S^{(int-n)}_t &:=& S_0\exp\le((r-d)t +  \int_0^t \sqrt{Z^{(n)}_s}\td X^{(n)}_s - \int_0^t\psi_n\le(-\cu\sqrt{Z^{(n)}_s}\ri)\td s\ri),\\ 
S^{(tc-n)}_t &:=& S_0\exp\le((r-d)t +  X^{(n)}\le(V^{(n)}_t\ri) - \psi_n(-\cu) V^{(n)}_t\ri),\\
&\phantom{=}& \text{where}\q V^{(n)}_t = \int_0^tZ^{(n)}_s \td s.
\end{eqnarray*}
The processes $S^{(int-n)}$ and $S^{(tc-n)}$ are in law equal to exponential L\'{e}vy processes:
\begin{prop}
For $n\in\mathbb N$, $\log S^{(int-n)}$ and $\log S^{(tc-n)}$ 
are in law equal
to regime-switching L\'{e}vy processes of the form \eqref{eq:Process}.
\end{prop}
\begin{pr}
Note first that 
since $X^{(n)}=(X^{(n)}_t)_{t\ge 0}$ is a L\'{e}vy process 
with DPH jumps it is of the form 
\begin{equation}
X^{(n)}_t = \mu^{(n)} t + \sigma^{(n)} W_t + J^{(n)}_t\qquad \text{with}\qquad J^{(n)}_t = \sum_{i=1}^{M^{(n)}_t} U_i,
\end{equation}
where $\mu^{(n)}, \sigma^{(n)}$ are constants, $M^{(n)}=(M^{(n)}_t)_{t\ge 0}$ Poisson processes with jump-rates $\lambda^{(n)}$, and $U_i$ are i.i.d. random variables 
following a DPH distribution.
Then it is clear that $X^{(int-n)} = \log(S^{(int-n)}/S_0)$ is in law equal to the process $\WT X^{(int-n)}=(\WT X^{(int-n)}_t)_{t\ge 0}$ given by
\begin{eqnarray*}
\WT X^{(int-n)}_t &=& \int_0^t\le(r-d+\mu^{(n)}\sqrt{Z^{(n)}_s} - \psi^{(n)}(-\cu \sqrt{Z^{(n)}_s})\ri)\td s + 
\int_0^t\sigma^{(n)}\sqrt{Z^{(n)}_s}\td W_s \\
&\phantom{=}& + \sum_{j=1}^{N^{(n)}}\int_0^tI_{\{Z^{(n)}_s=x^{(n)}_j\}}\td \WT J^{(n,j)}_s,
\end{eqnarray*}
where $\WT J^{(n,j)}, j=1, \ldots, N^{(n)}$, 
are independent compound Poisson processes, that are 
in law equal to the processes $J^{(n,j)}=(J^{(n,j)}_t)_{t\ge 0}$ with
$J^{(n,j)}_t=\sum_{i=1}^{N^{(n)}_t} \sqrt{x^{(n)}_j}U_i$, respectively.
Since, for any constant $c\neq 0$, $c U_i$ follows a DPH distribution, $\WT X^{(int-n)}$ is a regime-switching L\'{e}vy process of the form \eqref{eq:Process}.

As a consequence of the scaling property of Brownian motion 
it follows that $X^{(tc-n)}=\log (S^{(tc-n)}/S_0)$ 
is in law equal to the process 
$\WT X^{(tc-n)}=(\WT X^{(tc-n)}_t)_{t\ge 0}$ given by
\begin{eqnarray*}
\WT X^{(tc-n)}_t &=& \int_0^t\le(r-d + 
\le[\mu^{(n)} - \psi^{(n)}(-\cu)\ri]Z^{(n)}_s\ri)\td s 
+ \int_0^t\sigma^{(n)}\sqrt{Z^{(n)}_s}\td W_s\\ 
&\phantom{=}& + \sum_{j=1}^{N^{(n)}}
\int_0^tI_{\{Z^{(n)}_s=x^{(n)}_j\}}\td \WH J^{(n,j)}_s,
\end{eqnarray*}
where $\WH J^{(n,j)}, j=1, \ldots, N^{(n)}$, 
are independent compound Poisson processes, that are 
in law equal to the processes $J^{(n,j)}=(J^{(n,j)}_t)_{t\ge 0}$ with
$J^{(n,j)}_t=\sum_{i=1}^{M^{(n,j)}_t} U_i$, respectively, 
where $M^{(n,j)}$ is a Poisson process with jump-rate 
$\lambda \cdot x^{(n)}_j$. Here we used that, conditional on 
$Z^{(n)}$, the process $X^{(tc-n)}$ has independent increments,
so that the law of $X^{(tc-n)}$ conditional on $Z^{(n)}$ 
is determined by the conditional characteristic functions 
of $X^{(tc-n)}_t$, 
$t\ge0$. A straightforward calculation verifies that 
the conditional characteristic functions 
of $X^{(tc-n)}_t$ and $\WT X^{(tc-n)}_t$ are equal for $t\ge 0$.
\end{pr}

To establish the convergence in law of $(S^{(int-n)})_{n\in\NN}$ and
$(S^{(tc-n)})_{n\in\NN}$ we will
first show convergence of the finite-dimensional
distributions.\footnote{A sequence of processes 
$(Y_n)_{n\in\NN}$
converges in finite dimensional distribution to the process $Y=\{Y_t\}_{t\in[0,T]}$, if, for any
partition $t_1 < \ldots <  t_m$ of $[0,T]$, $\PP(Y^{(n)}_{t_1}\leq x_1, \ldots, Y^{(n)}_{t_m}\leq x_m)\to 
\PP(Y_{t_1}\leq x_1, \ldots, Y_{t_m}\leq x_m).$  
We will denote this convergence by $Y_n\stackrel{fidis}{\Rightarrow} Y$.}

\begin{prop}\label{prop:fidis}
Assume that $X^{(n)}\stackrel{\mathcal L}{\Rightarrow} X$ 
and $Z^{(n)}\stackrel{\mathcal L}{\Rightarrow} v$ 
as $n$ tends to infinity. 
Then the following holds true:

(a) $(Z^{(n)}, S^{(int-n)})_n \stackrel{fidis}{\Rightarrow} (v, S)$
where $S$ is the model given in \eqref{eq:Model2}.

(b) $(V^{(n)}, S^{(tc-n)})_n \stackrel{fidis}{\Rightarrow} (V, S)$ 
where $S$ is the time-change model given in \eqref{eq:Model_Gang2}.

\end{prop}

%
\noindent{\bf Remark.} The convergence of European put option prices (and hence, by put-call parity, also of European call option prices)
under the approximating models to those under the limiting models is a direct consequence of the convergence in finite dimensional distributions. 
To establish the convergence of path-dependent option prices such as barrier option
prices it is required to prove that the approximating models converge in law.

\begin{pr}
(b)  To prove the convergence of the finite dimensional distributions
it suffices, in view of the Markov property, to show that, for each fixed $t\in[0,T]$, the characteristic functions $\chi_n$ of $(V^{(n)}_t, X^{(n)}(V^{(n)}_t))$
converge point-wise to the characteristic function $\chi$ 
of $(V_t,X(V_t))$ as $n$ tends to infinity. By conditioning and using the independence of $V^{(n)}$ from $X^{(n)}$ and of $V$ from $X$ we find that 
\begin{eqnarray*}
\chi_n(u,v) &=& 
\EE_{x,i}\le[\exp\le\{(\cu u + \psi_n(v))V^{(n)}_t\ri\}\ri],
\qquad \chi(u,v) = 
\EE_{x,i}[\exp((\cu u + \psi(v))V_t)].
\end{eqnarray*}
Since 
$v^{(n)}$ converges in law to $v$ in the Skorokhod topology,
the Skorokhod 
representation theorem implies that on some probability space $Z^{(n)}\to v$, almost surely, with the convergence with respect to the Skorokhod metric.
Since, for any $t\in[0,T]$, the map $i_t: D_{\mathbb R}[0,T]\to\mathbb R$ given by $i_t: x\mapsto \int_0^t x(s)\td s$, is continuous in the Skorokhod topology, we deduce that $V^{(n)}_t \to V_t$ almost surely. In particular, $\chi_n$ converges point-wise to $\chi$.

The proof of (a) is similar and omitted.
\end{pr}

The next result concerns the convergence in law of 
the sequences
$(S^{(int-n)})_{n\in\NN}$ and $(S^{(tc-n)})_{n\in\NN}$:

\begin{thm}
The following statements hold true:

(a) 
Assume that $(v,S)$ is a Feller process, 
where $S$ is the model given in \eqref{eq:Model2}, that 
$Z^{(n)}$ satisfies the conditions in Theorem \ref{thm:conv},
and that the characteristics of $X^{(n)}$ satisfy conditions \eqref{cnsn}
and \eqref{eq:nunnu}. Then, as $n\to\infty$, 
$$
S^{(int-n)}\stackrel{\mathcal L}{\Rightarrow} S.
$$

(b) Assume that $X^{(n)}\stackrel{\mathcal L}{\Rightarrow} X$ 
and $Z^{(n)}\stackrel{\mathcal L}{\Rightarrow} v$
as $n\to\infty$. Then it holds that
$$
S^{(tc-n)}\stackrel{\mathcal L}{\Rightarrow} S,
$$
where $S$ is the time-change model given in \eqref{eq:Model_Gang2}.
\end{thm}
{\bf Remark.} The convergence in law stated above carries over to the convergence 
of barrier option prices under the respective models, if the boundaries are continuity points of the limiting model. For instance, if we denote by $\tau_A=\inf\{t\ge0: S_t\notin A\}$ the first time that $S$ leaves the set $A:=[\ell,u]$, and
$\PP(S_T\in\{\ell,u\})=0$, then, for any bounded continuous pay-off 
functions 
$g, h: \mathbb R_+\to\mathbb R$, we have that as $n\to\infty$ 
the double knock-out option and rebate option prices under the approximating models converge to those under the limiting model:
\begin{eqnarray*}
\EE_x\le[
g\le(S^{(n)}_{T}\ri)I_{\{\tau_A^{(n)}>T\}}\ri]
&\longrightarrow&
\EE_x\le[
g(S_{T})I_{\{\tau_A>T\}}\ri],\\
\EE_x\le[\te{-r\tau_A^{(n)}}
h\le(S^{(n)}_{\tau^{(n)}_A}\ri)I_{\{\tau_A^{(n)}\leq T\}}\ri]
&\longrightarrow&
\EE_x\le[\te{-r\tau_A} 
h(S_{\tau_A})I_{\{\tau_A \leq T\}}\ri], 
\end{eqnarray*}
where $S^{(n)}$ denotes $S^{(int-n)}$ or $S^{(tc-n)}$. 
A proof of this result was given in \cite{Pistorius_MarkovBarrier:2009}.

\begin{pr}
In view of Proposition \ref{prop:fidis}, it suffices\footnote{See e.g. Theorem 3.7.8 in Ethier and Kurtz \cite{Kurtz} for a proof of this well known fact.} to verify that the sequences 
$(S^{(int-n)})_{n\in\NN}$ and $(S^{(tc-n)})_{n\in\NN}$ are relatively compact 
in $D_{\mathbb R}[0,T]$.

(a) We will establish relative compactness of 
the sequence $(X^{(int-n)})_{n\in\NN}=(\log S^{(int-n)})_{n\in\NN}$.
Let 
$X'=\log S$.
It is straightforward to check that the set of functions $f:\mathbb R_+\times\mathbb R_+\to\mathbb R$ of the form $f(x,v)=g(x) h(v)$ with $h$ in the core of $\mc G$, the infinitesimal generator of $v$, and with 
$g\in C_c^\infty(\mathbb R)$\footnote{$C_c^\i(\mathbb R_+)$ 
denotes the set 
of infinitely differentiable functions with compact support contained in $\mathbb R_+$.} is dense in $C_0(\mathbb R_+\times\mathbb R)$
and is contained in the domain $\mc D(\mathcal L')$ of  the infinitesimal
generator $\mathcal L'$ of $(v,X')$. Furthermore, 
$\mathcal L'$ acts on such $f$ as
\begin{eqnarray*}
\mc L'f(x,v) &=& \frac{1}{2}vh(v)\sigma^2 g''(x) + 
\le[(r-d) + c\sqrt{v} 
- \psi(-\cu\sqrt{v})\ri]h(v)g'(x)\\
&+& h(v)\int_{\mathbb R} \le[g(x + z\sqrt{v}) - g(x) - z\sqrt{v}g'(x)I_{\{|z|\leq1\}} \ri]\nu(\td z) + g(x)\mc G h(v),\qquad x\in\mathbb R, v> 0,
\end{eqnarray*}
since by construction the stochastic integral 
$\left\{\int_0^t\sqrt{v_s}\td X_s\right\}_{t\geq0}$
jumps if and only if the L\'evy process 
$X$
jumps and, if the jump occurs at time 
$t$,
the quotient of the jump sizes equals
$\sqrt{v_t}$.
On the other hand, the infinitesimal generator $\mc L^{(n)}$ 
of the regime-switching processes $(Z^{(n)}, X^{(int-n)})$ 
acts on $f(x,v)=g(x) h(v)$ as
\begin{eqnarray*}
\mc L^{(n)}f(x,v) &=& \frac{1}{2}v\sigma^2_n h(v) g''(x) + 
\le[(r-d)+ c_n\sqrt{v} -\psi_n(-\cu\sqrt{v})\ri]h(v)g'(x)\\
&+& h(v)\int_{\mathbb R} \le[g(x + z\sqrt{v}) - g(x) - z\sqrt{v}g'(x)I_{\{|z|\leq1\}}\ri]\nu_n(\td z) + g(x) Q^{(n)} h(v),
\qquad x\in\mathbb R, v\in E^{0(n)}.
\end{eqnarray*}
 $\mc L^{(n)}f$ converges to $\mc L'f$ uniformly as $n\to\infty$:
\begin{equation}
\epsilon'_n(f) := \sup_{x\in\mathbb R, v\in E^{0(n)}}|\mc L'f(x,v) - \mc L^{(n)}f(x,v)| \to 0.
\end{equation}
To see why this is true note that the triangle inequality and integration by parts imply that
\begin{eqnarray}\nn
\lefteqn{\epsilon'_n(f)\leq \|\mc G h - Q^{(n)} h\|_n \|g\|_\infty  
+ C_1 |\ovl\nu(a) - \ovl\nu_n(a)| + C_2\max\{|\sigma_n^2 - \sigma^2|,|c_n-c|\}}
\\ \nn &+& 
C_3\le\{\int_{a}^\infty|\ovl\nu(z) - \ovl\nu_n(z)|\td z+ 
\int_{-\infty}^{-a}|\unl\nu(z) - \unl\nu_n(z)|\td z\ri\}\\ 
&+& 
C_4\le\{\int_{(0,a)}z^2|\ovl\nu(z) - \ovl\nu_n(z)|\td z + \int_{(-a,0)}z^2|\unl\nu(z) - \unl\nu_n(z)|\td z\ri\}, \end{eqnarray}
where $a$ is a continuity point of the measures $\nu(\td x)$ and $\nu(-\td x)$ and
$C_1, \ldots, C_4$, are certain finite constants independent of $n$, and we denoted $\|f\|_n=\sup_{x\in E^{0(n)}}|f(x)|$ and 
$\|f\|_\infty = \sup_{x\in\mathbb R}|f(x)|$. In view of the conditions 
\eqref{cnsn} and \eqref{eq:nunnu} $\epsilon_n'(f)$ tends to zero as $n$ tends to infinity.
Corollary 4.8.6 in Ethier and Kurtz \cite{Kurtz} implies then that 
$(Z^{(n)}, X^{(int-n)})_{n\in\NN}$ and hence $(Z^{(n)}, S^{(int-n)})_{n\in\NN}$ 
is relatively compact in $D_{\mathbb R^2}[0,T]$.

(b) Denote by $\WT X^{(n)}=\{\WT X^{(n)}_t\}_{t\ge 0}$ and 
$\WT X=\{\WT X_t\}_{t\ge 0}$ the L\'{e}vy processes 
given by 
$$
\WT X^{(n)}_t = 
X^{(n)}_t - \psi^{(n)}(-\cu) t
\q\text{and}\q
\WT X_t = X_t - \psi(-\cu) t.
$$
We will verify\footnote{The proof draws on and combines a number of 
results from the theory of weak convergence of probability measures
that can be found in Ethier and Kurtz \cite[Chapters 3, 6]{Kurtz}} the relative compactness
of the sequence $(Y^{(n)})_{n\in\NN}$
with $Y^{(n)}_t = \WT X^{(n)}(V^{(n)}_t)$. 
In 
view of the Skorokhod embedding theorem
and the convergence in law of $(Z^{(n)},\WT X^{(n)})$ 
to $(v,\WT X)$, we may and shall assume that 
$(Z^{(n)},\WT X^{(n)})_{n\in\NN}$ and $(v,\WT X)$ are defined on the same 
probability space $(\Omega',\mc F',\PP')$, and that, $\PP'$-almost surely,
$(Z^{(n)}, X^{(n)})\to (v, X)$ with respect to the Skorokhod metric. 
Fix an $\omega\in\Omega'$ for which this convergence holds true.

Observe that, for any $U>0$, $\ovl z = \sup_{n} Z^{(n)}_U(\omega)$ is finite, as 
$v_U(\omega)$ is finite and $|Z^{(n)}_U(\omega) - v_U(\omega|\to 0$ as $n$ tends to infinity.
For $x\in D_{\mathbb R}[0,U]$, $\delta>0$, $U>0$, denote
by $w'(x,\delta,U)$ the modulus of continuity
$$
w'(x,\delta,U) = \inf_{\{t_i\}}\max_i \sup_{s,t\in[t_{i-1},t_i)}|x(s)-x(t)|,
$$
where $\{t_i\}$ ranges over all partitions  $0=t_0<t_1 < \ldots < t_{n-1} < U\leq t_n$ with $\min_{1\leq i\leq n}|t_i-t_{i-1}|>\delta$.
Note that 
$w'(x,\delta,U)$
is non-decreasing in 
$\delta$
and
$U$.
Therefore it is straightforward to check that
\begin{eqnarray}
w_n'(U) := w'(Y^{(n)}(\omega),\delta, U) &\leq& w'(\WT X^{(n)}(\omega),\delta\ovl z, U\ovl z).
\end{eqnarray}
Furthermore, observe that
\begin{eqnarray}
\mathcal Y_n(\omega) :=\{Y^{(n)}_s(\omega): s\leq U\} &\subset& \{\WT X^{(n)}_s(\omega): s\leq U\ovl z\}.
\end{eqnarray}
Since $\{\WT X^{(n)}(\omega)\}_{n\in\NN}$ is convergent in $D_{\mathbb R}[0,\infty)$, 
it follows that 
(I) for every rational $t\in [0,U]$ there exists 
a compact set $C_t\subset\mathbb R$ such that 
$Y^{(n)}_t(\omega)\in C_t$ for all $n$ and (II) for every $U>0$, 
$\lim_{\d\to0}\sup_n w'_n(U)=0$
and as a consequence\footnote{Both applications follows from the fact that conditions
(I) and (II) are necessary and sufficient for the 
relative compactness of $(Y^{(n)})_{n\in\NN}$ 
(Ethier and Kurtz \cite[Theorem 3.6.3]{Kurtz}).} $\mathcal Y_n$ is relatively compact in $D_{\mathbb R}[0,\infty)$. 
\end{pr}

\section{European and volatility derivatives}
\label{sec:EuropVol}

\subsection{Call and put options} 
\label{subsec:Eur}

We first turn to the valuation of a call option price. 
In the model under consideration a closed form expression is available,
in terms of the original parameters, for the Fourier transform $\FT{c}_T$ in log-strike $k=\log K$ of the call prices $C_T(K)$ with maturity $T$, 
$$
\FT{c}_T(\xi) = \int_{\mathbb R} \te{\mrm i\xi k}C_T(\te{k})\td k\quad\text{where}
\quad \Im(\xi)<0.
$$

\begin{prop} 
\label{Prop:FourTrnsformVanilla}
Define for any 
$\xi\in\CC\backslash\{0,\cu\}$,
$x\in\RR$
and
$j\in E^0$
the value 
$D(\xi,x,j)$
is defined 
by the formula
\begin{equation}
\label{eq:DefD}
D(\xi,x,j):= \frac{\te{(1+\cu\xi)x}}{{\cu\xi - \xi^2}}
\cdot\le[\exp\big\{T(K(1+\cu\xi) - \Lambda_D)\big\}\mbf 1\ri](j).
\end{equation}
Then if 
$\Im(\xi)<0$ 
it holds that
\begin{equation*}
\FT{c}_T(\xi) = D(\xi,x,j)
\end{equation*}
where $x=\log S_0$ is the log-price at the current time and $Z_0=j$ the initial 
level of the volatility. 
\end{prop}
{\bf Remarks.} (i) The call option price can now be calculated using the method 
described in Carr-Madan~\cite{CarrMadan} 
by evaluating the
integral
\begin{eqnarray}
C_T(K)
& = & \frac{\exp(-\alpha k)}{2\pi}\int_{-\infty}^\infty e^{-\cu sk}\FT{c}_T(s-\cu
\alpha)\td s \nonumber\\
& = & \frac{\exp(-\alpha k)}{\pi}\int_{0}^\infty \Re\left[e^{-\cu sk}D(s-\cu\alpha,\log
S_0,Z_0)\right]\td s,
\label{eq:Callprice} 
\end{eqnarray}
for $k=\log(K)$
and any strictly positive 
$\alpha$.
The integral in~\eqref{eq:Callprice} can be
approximated efficiently by a finite sum using the FFT algorithm
(see~\cite{CarrMadan}). Since in our model we have an explicit
formula for the transform $\FT{c}_T(s)$ given
by~\eqref{eq:DefD}, 
the pricing of European call options is
immediate. 

(ii) A simple calculation 
shows that the put option price
$P_T(K)=\EE_{x,j}\left[(B_T^{D})^{-1} (K-S_T)^+\right]$
can be expressed in terms of the 
formula
for
$D(\xi,x,j)$
in~\eqref{eq:DefD}
and any strictly negative
constant 
$\alpha$
in the following way:
\begin{eqnarray*}
P_T(K)
& = & \frac{\exp(-\alpha k)}{\pi}\int_{0}^\infty \Re\left[e^{-\cu sk}D(s-\cu\alpha,\log
S_0,Z_0)\right]\td s,
\quad
\text{where}\quad k=\log(K).
\end{eqnarray*}


\begin{pr} To find the European call option price
$C_T(K)=\EE_{x,j}\left[(B_T^{D})^{-1} (S_T-K)^+\right]$
in model~\eqref{eq:Process}
we first need to find the Fourier transform 
in the log-strike
$k=\log(K)$
of the function
$$c_T(k)=\exp(\alpha k)\EE_{x,j}\left[(B_T^{D})^{-1} (S_T-\exp(k))^+\right],$$
where 
$\alpha$
is some strictly positive constant. 
Fubini's theorem
and the form of the characteristic function~\eqref{eq:ch}
imply the following for 
$\xi=v-\cu \alpha$:
\begin{eqnarray*}
\FT{c}_T(\xi) & = &
\int_\RR \exp((\cu v+\alpha) k)
\EE_{x,j}\left[(B_T^{D})^{-1} (S_T-\exp(k))^+\right] \td k \nonumber \\
& = &
\EE_{x,j}\left[(B_T^{D})^{-1} \int_\RR \exp((\cu v+\alpha) k)
(S_T-\exp(k))^+\td k\right]\nonumber \\
& = &
\EE_{x,j}\left[(B_T^{D})^{-1} \exp((1+\alpha+\cu v)X_T)\right]/ (\alpha^2+\alpha-v^2+\cu(2\alpha+1)v) \nonumber\\
& = & 
\frac{\te{x(1+\alpha+\cu v)}}{\alpha^2+\alpha-v^2+\cu (2\alpha+1)v}
\le[\exp(T(K(1+\alpha+\cu v)-\Lambda_D))\1\ri](j).
\end{eqnarray*}
This concludes the proof.
\end{pr}

\subsection{Implied volatility at extreme strikes}
\label{subsec:IV}

The \textit{implied volatility}
$\sigma_{x,i}(K,T)$
for a given strike 
$K$
and maturity
$T$
is uniquely defined by the identity
\begin{eqnarray}
\label{eq:IVolDef}
C^{\text{BS}}(S_0,K,T,\sigma_{x,i}(K,T))=\EE_{x,i}\left[(B_T^{D})^{-1} (S_T-K)^+\right],
\end{eqnarray}
where
$C^{\text{BS}}(S_0,K,T,\sigma)$
is the Black-Scholes formula
and
$S_0=\exp(x)$.
The results in Lee~\cite{Lee}
and refinements in Benaim and Friz \cite{BenaimFritz2}
imply that in model~\eqref{eq:Process}
the slope of the volatility smile is uniquely determined
by the 
quantities
$\alpha^\pm_i$,
$i=1,\ldots,n$,
defined in~\eqref{eq:SmallestEigVal}.
In the particular case where the distribution of jumps is double
exponential, 
$\alpha^\pm_i$
are in fact the reciprocals of
mean-jump sizes in model~\eqref{eq:Process}.

In order to state the precise result, define 
\begin{eqnarray*}
q_+ &:=& \sup\left\{u: \EE_{x,i}\left[S_T^{1+u}\right] < \infty\quad\text{for all}\quad i\in E^0\right\},\\
q_- &:=& \sup\left\{u: \EE_{x,i}\left[S_T^{-u}\right] < \infty\quad\text{for all}\quad i\in E^0\right\}.
\end{eqnarray*}
If the chain $Z$ is irreducible, the quantities $q_\pm$ can be identified explicitly to be equal to
\begin{eqnarray}\label{qp}
q_+ &=& \min\{\alpha_i^+ - 1\>:\>i\in\{1,\ldots,N_0\}\>\> \&\>\> p_i \lambda_i>0 \},\\ 
q_- &=& \min\{\alpha_i^-\>:\>i\in\{1,\ldots,N_0\}\>\>\&\>\>(1-p_i) \lambda_i>0 \}.\label{qm}
\end{eqnarray}
As noted above
the quantities 
$q_+$
and
$q_-$
depend only on the mean-jump sizes 
of the compound Poisson processes in model~\eqref{eq:Process}.

Denote the forward price by
$F_T:=\EE_{x,i}[S_T]$.
Then the asymptotic behaviour for the implied volatility 
is described as follows:
\begin{prop} Suppose that $Z$ is irreducible. 
For $T>0$ and $K>0$ and with $q_\pm$ given 
in \eqref{qp}--\eqref{qm}, it holds that 
\begin{eqnarray*}
\lim_{K\to\infty}\frac{T\sigma_{x,i}(K,T)^2}{\log(K/F_T)}  &=&  2-4\left(\sqrt{q_+^2+q_+}-q_+\right),\\
\lim_{K\to0}\frac{T\sigma_{x,i}(K,T)^2}{\lvert \log(K/F_T)\rvert}  &=&  2-4\left(\sqrt{q_-^2+q_-}-q_-\right).
\end{eqnarray*}
\end{prop}
{\bf Remark.}  Note that, if the Markov chain $Z$ is irreducible,  the asymptotic slope of the implied volatility 
smile for large and small strikes depends neither on the spot $S_0=\te{x}$ nor 
on the starting volatility regime $i$.

\subsection{Forward starting options and the forward smile}
A forward starting call option is a call option whose strike is fixed at a later date as a proportion of the value of the underlying at that moment.
More precisely, the pay-off of a $T_1$-forward starting call option 
at maturity $T_2>T_1$ is given by
$$
(S_{T_2} - \kappa S_{T_1})^+,\qquad\qquad \kappa\in\RR_+.
$$
Denote the current value of this forward starting option by $F_{T_1,T_2}(\kappa)$
and let 
$$
\FT{F}_{T_1,T_2}(\xi) = \int_{\mathbb R} \te{\mrm i\xi k}F_{T_1,T_2}(\te{k})\td k,\quad\text{where}
\quad \Im(\xi)<0,
$$
be its Fourier transform in the forward log-strike 
$k=\log\kappa$.

\begin{prop} 
\label{prop:FrowardStart}
For $\xi$ with $\Im(\xi)<0$ it holds that
\begin{equation}
\FT{F}_{T_1,T_2}(\xi) = 
\frac{\te{(1+\cu\xi)x}}{{\cu\xi - \xi^2}}
\cdot 
\le[\exp(T_1(Q-\Lambda_F))\exp\big\{(T_2-T_1)(K(1+\cu\xi) - \Lambda_D)\big\}\mbf 1\ri](j),%
\end{equation}
where $x=\log S_0$ is the log-spot price  and $Z_0=j$ the initial 
level of the volatility.
\end{prop}

\noindent {\bf Remark.} An inversion formula, analogous to the one
in~\eqref{eq:Callprice}, can be used to obtain the value
$F_{T_1,T_2}(\kappa)$
from Proposition~\ref{prop:FrowardStart}.

%


\begin{pr}
The price of a $T_1$-forward starting option 
is given by the expression
$$
F_{T_1, T_2}(\kappa) = \EE_{x,i}\left[(B^D_{T_2})^{-1} (S_{T_2} - \kappa S_{T_1})^+\right], \quad \kappa\in\RR_+.
$$
The process 
$(X,Z)$
is Markov and therefore, by conditioning on the 
$\sigma$-algebra generated by the process up to time
$T_1$, employing the form \eqref{eq:ch} 
of the characteristic function of $X_T$
and using the spatial homogeneity of the log-price $X_t=\log S_t$
in our model, we obtain the following expression for the price of the forward starting option
\begin{eqnarray}
F_{T_1, T_2}(\kappa) & = & \EE_{x,i}\left[(B^D_{T_2})^{-1} (S_{T_2} - \kappa S_{T_1})^+\right]\\ 
&=& \EE_{x,i}\left[\frac{S_{T_1}}{B^D_{T_1}}
\EE_{0,Z_{T_1}}\left[(B^D_{T_2-T_1})^{-1}(S_{T_2-T_1}-\kappa)^+\right]  \right] \nonumber \\
&=&  \sum_{j\in E^0} \EE_{x,i}\left[\frac{S_{T_1}}{B^D_{T_1}}I_{\{Z_{T_1}=j\}}\right]
\EE_{0,j}[(B^D_{T_2-T_1})^{-1}(S_{T_2-T_1}-\kappa)^+]\nonumber \\
& = & S_0 \sum_{j\in E^0} e_i'\exp(T(K(-\cu)-\Lambda_D))e_j \EE_{0,j}[(B^D_{T_2-T_1})^{-1}(S_{T_2-T_1}-\kappa)^+]\nonumber \\
& = & S_0 e_i'\exp(T(K(-\cu)-\Lambda_D))C_{T_2-T_1}(\kappa;1), 
\label{eq:FwdStart}
\end{eqnarray}
where
$C_{T_2-T_1}(\kappa;1)$
is a vector (of call option prices)
with 
$j$-th component equal to
$\EE_{0,j}[(B^D_{T_2-T_1})^{-1}(S_{T_2-T_1}-\kappa)^+]$.
The martingale condition in~\eqref{mc2}
and Proposition~\ref{Prop:FourTrnsformVanilla}
conclude the proof.
\end{pr}

\noindent {\bf Remarks.} (i)  A quantity of great interest in the derivatives
markets 
is the \textit{forward implied volatility}
$\sigma^{fw}_{x,i}(S_T,\kappa,T)$
at a future time $T$ implied by the model.
It is defined as the unique solution to the equation
\begin{equation}
\label{eq:FVolDef}
C^{\text{BS}}(S_{T_1},\kappa S_{T_1},T_2-T_1,\sigma^{fw}_{x,i}(S_{T_1},\kappa,T_1)) 
= 
\EE_{x,i}\left[\frac{B_{T_1}^{D}}{B_{T_2}^{D}} (S_{T_2}-\kappa S_{T_1})^+\bigg|S_{T_1}\right],
\end{equation}
where the left-hand side denotes the Black-Scholes formula
with strike
$\kappa S_{T_1}$
and spot
$S_{T_1}$.
The reason for the importance of the forward implied volatility
$\sigma^{fw}_{x,i}(S_T,\kappa,T)$
lies in the problem of hedging of exotic derivatives using vanilla options.
If at a future time
$T$
the spot trades at the level 
$S_T$,
then the trader needs to know where, according to the model, 
would the vanilla surface be trading at.
This is of particular importance when hedging a barrier contract
that knocks out at the level
$S_T$,
because conditional on this event
the trader is left with a portfolio of vanilla 
options that was created as a semi-static hedge for 
the exotic derivative.

(ii) In the model given by~\eqref{eq:DomesticBond}
we can compute the right-hand side of~\eqref{eq:FVolDef}.
In view of the Markov property of the process $(X,Z)$, this equation 
is equivalent to 
\begin{eqnarray*}
C^{\text{BS}}(S_{T_1},\kappa S_{T_1},T_2-T_1,\sigma^{fw}_{x,i}(S_{T_1},\kappa,T_1)) 
& = & 
S_{T_1}\EE_{x,i}\left[
\EE_{0,Z_{T_1}}\left[(B^D_{T_2-T_1})^{-1}(S_{T_2-T_1}-\kappa)^+\right] \big\lvert\>
S_{T_1} \right] \nonumber \\
&=& S_{T_1} \sum_{j\in E^0} \PP_{x,i}\left[ Z_{T_1}=j\Big\lvert\> S_{T_1}\right]
\EE_{0,j}[(B^D_{T_2-T_1})^{-1}(S_{T_2-T_1}-\kappa)^+]\nonumber \\
& = & S_{T_1} f^{x,i}(X_{T_1},T_1)' C_{T_2-T_1}(\kappa,1), 
\end{eqnarray*}
where the coordinates of the vector
$f^{x,i}(y,T)$
are defined by
\begin{eqnarray}
\label{eq:DefOfVectf}
f^{x,i}_j(y,T):= 
\PP_{x,i}\left[Z_{T}=j\Big\lvert\> X_{T}=y\right]
\end{eqnarray}
and
$C_{T_2-T_1}(\kappa;1)$
is as defined in the line following equation~\eqref{eq:FwdStart}.

(iii) The vector 
$C_{T_2-T_1}(\kappa;1)$
can be computed by formula~\eqref{eq:Callprice}
and, in the light of definition~\eqref{eq:DefOfVectf},
Proposition~\ref{prop:ExisteceOfDensity}
and formulae~\eqref{eq:ForwardSmileVector} and~\eqref{eq:DensityOfX} for
$q^{x,i}_T(y,j)$
and
$q^{x,i}_T(y)$,
it follows that
$$
f^{x,i}_j(y,T)q^{x,i}_T(y) = q^{x,i}_T(y,j).
$$
This yields the quantity in~\eqref{eq:DefOfVectf}
and hence a formula for the forward implied volatility
in our model.

\subsection{Volatility derivatives}

An option on the realized variance is a
derivative security that delivers
$\phi(\Sigma_T)$
at expiry
$T$,
where
$\phi:\RR_+\to\RR$
is some measurable payoff function
and
$\Sigma_T$
is the quadratic variation
up to time
$T$
of the process
$\log S = X$.
More formally, for
a refining sequence of partitions\footnote{
The sets
$\Pi_n=\{t_0^n, t_1^n,\ldots,t_n^n\}$,
$n\in\NN$,
consist of increasing sequences of times
such that
$t_0^n=0$,
$t_n^n=T$,
$\Pi_n\subset\Pi_{n+1}$
for
all
$n\in\NN$
and
$\lim_{n\to\infty}\max\{|t_i^n-t_{i-1}^n|:i=1,\ldots,n\}=0$.}
$(\Pi_n)_{n\in\NN}$
of the interval
$[0,T]$, $\Sigma_T$ is given by
\begin{eqnarray*}
\Sigma_T&:=&\lim_{n\to\infty}\sum_{t_i^n\in\Pi_n, i\geq1}\log\left(\frac{S_{t_i^n}}{S_{t^n_{i-1}}}\right)^2\\
\end{eqnarray*}
It is well-known that the sequence on the right-hand side converges in probability, uniformly on compact time intervals
(see Jacod \& Shiryaev~\cite{JacodShiryaev}, Theorem 4.47)
and the limit is given by
\begin{equation}\label{eq:sigma}
\Sigma_T = \int_0^T\sigma(Z_t)^2\td t + \sum_{i\in E^0}\sum_{t\leq T}I_{\{Z_t=i\}}(\Delta J_t^i)^2.
\end{equation}
where $\Delta J^i_t:=J^i_t-J^i_{t-}$. The process
$\{\Sigma_t\}_{t\geq0}$ is called the quadratic variation or
realized variance process of $X$, and its law is explicitly
characterised as follows:

\begin{prop}\label{pr:s}
(i) The process $\{(\Sigma_t, Z_t)\}_{t\ge 0}$ is a Markov process with
$$\Sigma_t= \int_0^t\sigma(Z_s)^2\td s + \sum_{i\in E^0}\int_0^tI_{\{Z_s=i\}}d \WT J_s^i,$$
where
$\WT J^i$,
$i\in E^0$,
is a compound Poisson process with intensity
$\lambda_i$
and positive jump sizes
$K_i$ with probability density
$$
g_i(x) = \frac{1}{2\sqrt{x}}\le[p_i\beta_i^+\te{\sqrt{x}B_i^+}(-B_i^+)\mbf 1 + (1-p_i)\beta_i^-\te{\sqrt{x}B_i^-}(-B_i^-)\mbf 1
\ri] I_{(0,\infty)}(x).
$$
(ii) The discounted Laplace transform of $\Sigma_t$ is given by
\begin{equation}\label{eq:charsigmaT}
\EE_{i}\left[\frac{\exp(- u\Sigma_t)}{B_t^D}\right]=
\le[\exp(t(K_{\Sigma}(u)-\Lambda_D))\1\ri](i), \qquad u>0,
\end{equation}
where $K_{\Sigma}(u) = Q + \Lambda_{\Sigma}(u)$
with $\Lambda_{\Sigma}(u)$ an $N_0\times N_0$ diagonal matrix  with $i$-th element given by
\begin{equation}
\psi_i^\Sigma(u):=- u\sigma_i^2+
\lambda_i\left(\EE\left[\exp(- u K_i)\right]-1\right),
\end{equation}
with
\begin{eqnarray*}
\EE\left[\exp(- u  K_i)\right]
& = & \sqrt{\frac{\pi}{u}}
\left(p_i \beta_i^+
\Phi\left(\frac{1}{\sqrt{2 u}}B_i^+\right)(-B_i^+)+
(1-p_i)\beta_i^-
\Phi\left(\frac{1}{\sqrt{2 u}}B_i^-\right)(-B_i^-)
\right)\mbf 1,
\end{eqnarray*}
where $\Phi(x):=\exp(x^2/2)\mathcal{N}\left(x\right)$,
with the cumulative normal distribution function $\mc N$.
\end{prop}
{\bf Remarks.} (i) As a given matrix $M$ in practice typically\footnote{This is the case since the set of all square matrices that do not
possess a diagonal decomposition is of co-dimension one in the space of all square matrices and therefore has Lebesgue measure zero.} admits a
spectral decomposition $M=U D U^{-1}$ where $D$ is a diagonal matrix,
$\Phi(M)$ can be evaluated by $\Phi(M)=U \Phi(D) U^{-1}$ where
$\Phi(D)$ is the diagonal matrix with $i$-th element $\Phi(D_{ii})$.

(ii) It is important to note that the realized variance process
$\Sigma$ does not possess exponential moments of any order. This follows directly  from the fact that the distribution of jumps
$g_i$ given in~\eqref{eq:JumpDistribSigma}
decays at the rate $e^{-c\sqrt{x}}$,
for some positive constant
$c$,
and implies that the left-hand side in formula~\eqref{eq:charsigmaT}
will be infinite for complex numbers
$u$ with negative imaginary part.

(iii) The expression~\eqref{eq:charsigmaT} for the discounted
Laplace transform can be employed to obtain explicit
results for the values of
volatility derivatives. The buyer of a swap on the realized variance
pays premiums at a certain rate (the swap rate) to receive at maturity
a pay-off $\phi(\Sigma_T)$ that is a  function $\phi$ of the realized variance
$\Sigma_T$, with as most common examples the volatility and the variance swap. In the case of a variance swap this function is linear ($\phi(x)=x/T$) whereas for a volatility swap it is a square root ($\phi(x) = \sqrt{x/T}$).
The swap rates are determined such that at initiation the value
of the swap is nil.

\begin{cor}\label{c:s} Suppose that $Z_0=j$.
Then the variance and volatility swap rates $\varsigma_{var}(T,j)$
and $\varsigma_{vol}(T,j)$ are given as follows:
\begin{eqnarray}
\label{eq:VolSwap}
\varsigma_{vol}(T,j) &=& \frac{1}{2\sqrt{\pi T}}
\int_0^\infty \big\{\left[\exp(T(Q-\Lambda_D))
-\exp(T(K_{\Sigma}(u)-\Lambda_D))\right]\1\big\} (j) \frac{\td u}{u^{3/2}},\\
\label{eq:varSwapPrice}
\varsigma_{var}(T,j) & = & \frac{1}{T}\int_0^T
\big[\exp\{t(Q - \Lambda_D)\}\Lambda_V
\exp\{(T-t)(Q - \Lambda_D)\}\1\big](j)\td t\\
& = & \frac{1}{Th}\left[\left\{\exp(T(Q-\Lambda_D))- \exp(T(K_{\Sigma}(h)-\Lambda_D))\right\}\1\right](j)+o(h), \ h\downarrow 0,
\label{eq:ld}
\end{eqnarray}
where $\Lambda_V$ is a $N_0\times N_0$ diagonal matrix with $i$-th element given by
\begin{equation}\label{eq:v}
V(i) =
\sigma_i^2 + 2\lambda_i\left(p_i(\beta_i^+)'(B_i^+)^{-2} +(1-p_i)(\beta_i^-)'(B_i^-)^{-2}\right)\1.
\end{equation}
\end{cor}
{\bf Remarks.} (i) The Laplace transform $\WH\sigma_{var}(q,j)$ of
$\sigma_{var}(\cdot,j):T\mapsto T\varsigma_{var}(T,j)$ is explicitly
given by
\begin{equation}
\WH\sigma_{var}(q,j) = \le[ (qI - \Lambda_D - Q)^{-1} \Lambda_V (qI - \Lambda_D - Q)^{-1}\mbf 1 \ri] (j).
\end{equation}

(ii) It is clear from the definition of
$K_\Sigma(u)$
that the integral in~\eqref{eq:VolSwap}
converges at the rate proportional to
$1/\sqrt{U}$
where
$U$
is an arbitrary upper bound used in the numerical
integration
in~\eqref{eq:VolSwap}.

\begin{pro}{Proposition \ref{pr:s}}
It is clear from the representation \eqref{eq:sigma},
that the increment $\Sigma_t- \Sigma_s$, for any $t>s\ge 0$,
satisfies the equation
\begin{equation}\label{sigmast}
\Sigma_t - \Sigma_s = \int_s^t\sigma^2(Z_u)\td u + \sum_{i\in E^0}
\int_s^t I_{\{Z_u = i\}}\td \WT J^i_u,
\end{equation}
where
$\WT J^i$,
$i\in E^0$,
is a compound Poisson process with intensity
$\lambda_i$
and positive jump sizes
$K_i$ distributed as $(U^i)^2$ where $U_i$ follows a
$DPH(p_i,\beta^+_i,B_i^+,\beta_i^-,B_i^-)$ distribution.
In particular, $K_i$ is distributed according to the density
\begin{eqnarray}
\label{eq:JumpDistribSigma}
g_i(x)  &= &\frac{1}{2\sqrt{x}} (f_i(\sqrt{x})+f_i(-\sqrt{x})),\quad x>0,
\end{eqnarray}
where the probability density function $f_i$ is given by
$$
f_i(x) = p_i (\beta_i^+)'\te{xB^+_i}(-B^+_i)\1
I_{(0,\infty)}(x) +
(1-p_i)(\beta^-_i)'\te{xB^-_i}(-B^-_i)\1
I_{(-\infty,0)}(x). 
$$
As the $\WT J^i$ have independent increments,
and $Z$ is a Markov chain, it directly follows from \eqref{sigmast}
that $(\Sigma_t,Z_t)$ is a Markov process, and moreover, a regime-switching subordinator. The form of the discounted Laplace transform can be derived as in Theorem \ref{thm:CharFun}.
\end{pro}

















































\begin{pro}{Corollary \ref{c:s}}
Employing the following representation for the square root
$$\sqrt{x}=\frac{1}{2\sqrt{\pi}} \int_0^\infty\left[1-\exp(-ux)\right]\frac{du}{u^{3/2}},\quad\text{for any}\quad x\geq0,$$
as well as the form of the discounted characteristic function given in~\eqref{eq:charsigmaT}
and Fubini's theorem yields that the volatility swap rate can be calculated
via a single one-dimensional integral
\begin{eqnarray*}
\EE_i\left[\frac{\sqrt{\mbox{$\frac{1}{T}$} \Sigma_T}}{B_T^D}\right] & = & \frac{1}{2\sqrt{\pi T}} \int_0^\infty e_i'\left[\exp(T(Q-\Lambda_D))
-\exp(T(K_{\Sigma}(u)-\Lambda_D))\right]\1\frac{\td u}{u^{3/2}}.
\end{eqnarray*}
The derivation of the variance swap rate formula \eqref{eq:varSwapPrice} rests on a conditioning argument. Indeed, by
conditioning on the sigma algebra
$\mc F^Z_T=\sigma(\{Z_t\}_{t\leq T})$ generated by $Z$
up to time $T$, it follows that
\begin{equation*}
\EE_{x,i}\le[\frac{1}{B_T^D}\le\{\int_0^t\sigma(Z_s)^2\td s + \sum_{i\in E^0}\int_0^tI_{\{Z_s=i\}}\td \WT J_s^i\ri\}\ri] = \sum_{j\in E^0}
\EE_{x,i}\le[\frac{1}{B_T^D}\int_0^T I_{\{Z_s = j\}}\td s\ri] w(j),
\end{equation*}
where $w(j) := \sigma^2(j) + \EE[\WT J^j_1]$.
From the definition of $\WT J^j$ it is easily checked that
$\EE[\WT J^j_1]$ is equal to $\lambda_j$ times the second moment
of the density $f_i$. One verifies by a straightforward calculation
that $w(j)$ is equal to $V(j)$ given in \eqref{eq:v}.
Furthermore, the Markov property of $Z$ applied at time $t$
yields that
$$
\EE_{x,i}\le[\frac{1}{B_T^D}\int_0^T I_{\{Z_s = j\}}\td s\ri]
= e_i' \exp(t(Q-\Lambda_D)) e_j' e_j \exp((T-t)(Q-\Lambda_D))\1.
$$
Equation \eqref{eq:varSwapPrice} follows then by an application of Fubini's theorem. Furthermore, Equation \eqref{eq:ld} follows by noting that
the expectation $\EE_{x,i}[\Sigma_T/B^D_T]$
can also be obtained by calculating the negative of the
derivative of the Laplace transform at zero:
\begin{equation*}
\EE_{x,i}\le[\frac{\Sigma_T}{B^D_T}\ri] =
 - \frac{\td}{\td u}\left\{\left[\exp(T(K_{\Sigma}(u)-\Lambda_D))\1\right](i)\right\}\Big\lvert_{u=0}.
\end{equation*}
\end{pro}

\section{First passage times for regime-switching processes}
\label{sec:AD}

\subsection{Three key matrices}
The characteristics of the process $(X,Z)$ can be summarised 
in terms of three matrices $Q_0$, $\Sigma$ and $V$ that will shortly 
be specified. Given those 
three matrices we will show how to reconstruct $(X,Z)$ in 
Section \ref{sec:FE}.

The matrices $Q_0$, $\Sigma$ and $V$ will be specified in nine-block matrices using
block-notation; the middle block of $Q_0$, $\Sigma$ and $V$ describes the rates 
of regime-switches, and the volatility and drift of the process in the different regimes, while the upper left and lower right block of $Q_0$ 
specify the distribution of up-ward and 
down-ward jumps in the different regimes, in terms of the (phase-type) generators.
More precisely, we define the three key matrices $Q_0, \Sigma$ and $V$, in block notation, by
\begin{eqnarray}\label{eq:Q0}
&Q_0 := \le(
\begin{array}{ccc}
 B^+ \q & b^+      & \q O    \\
 A^+ \q & Q-\Lambda_\lambda      & \q A^-  \\
 O   \q & b^-       & \q B^-  \\
\end{array}
\ri),&\\
&\Sigma := 
\le(
\begin{array}{ccc}
 O \q &   O    & \q  O  \\
 O  \q &  \Lambda_S      & \q O  \\
 O   \q &  O     & \q O  \\
\end{array}
\ri),
\q
V := 
\le(
\begin{array}{ccc}
 I \q &   O    & \q    O \\
 O  \q &  \Lambda_M      & \q  O  \\
 O   \q &   O     & \q - I  \\
\end{array}
\ri).&
\label{eq:SiV}
\end{eqnarray}
Here $Q$ is the generator matrix of the chain $Z$, and
 $\Lambda_\lambda$, $\Lambda_S$, $\Lambda_M$  denote
$N_0\times N_0$ diagonal 
matrices with elements 
\begin{equation}
\Lambda_\lambda(i,i) :=
\l_i,\q \Lambda_S(i,i) := \sigma(i)\q \text{and}\q 
\Lambda_M(i,i):=\mu(i).
\end{equation}
Further, $O$ 
and 
$I$
are zero and identity matrices of appropriate sizes 
such that 
$Q_0$,
$\Sigma$
and
$V$
are square matrices of the same dimension.
In block-notation $A^\pm$, $B^\pm$ and $b^\pm$ are given by 
\begin{equation}\label{Qa1}
A^\pm := \le(\begin{array}{ccc}
\l_1^\pm\b_1^{\pm\prime}&&\\
&\ddots&\\
&&\l_N^\pm\b_N^{\pm\prime}\\
\end{array}
\ri),\q B^\pm := \le(\begin{array}{ccc}
B_1^\pm&&\\
&\ddots&\\
&&B_N^\pm\\
\end{array}
\ri),\q b^\pm := \le(\begin{array}{ccc}
-B_1^{\pm}\1&&\\
&\ddots&\\
&&-B_N^{\pm}\1\\
\end{array}
\ri),
\end{equation}
where $\lambda_i^+:=\lambda_i p_i$ and $\lambda_i^-:=\lambda_i (1-p_i)$.

\smallskip

\no{\bf Remark.}  The matrix $Q_0$ is in fact the generator matrix of a Markov chain, as it has non-negative off-diagonal elements and zero row sums. 
We denote the state-space of this Markov chain by $E$. 
In the sequel we will frequently use the following partition of the set $E$:
\begin{equation}\label{eq:E}
E^\pm=\{i\in E:\Sigma_{ii}\neq 0\},  
\q E^+=\{i\in E:\Sigma_{ii}=0, V_{ii} > 0\}, 
\q E^- = \{i\in E: \Sigma_{ii}=0, V_{ii} < 0\}.
\end{equation}
Note further that 
$E^\pm$
can and will be identified with the state-space 
$E^0$
of the chain 
$Z$.
See Section~\ref{sec:FE} for further properties of the Markov 
chain
defined by the generator
$Q_0$. 
%

\subsection{Matrix Wiener-Hopf factorisation}
For a given vector of discount rates $h:E\to\CC$, the matrix Wiener-Hopf 
factorisation associates to the matrix 
$$Q_h := Q_0 - \Lambda_h,$$
where $\Lambda_h$ is a diagonal matrix with $i$-th diagonal element 
$\Lambda_h(i,i)=h(i)$, a quadruple of matrices 
which, as we will show below, characterises 
the distributions of the running maximum and minimum of $X$.
 
Let us briefly describe the sets of matrices of which this quadruple are elements. 
Denote by $\mathbb D(n)$ the set of $n\times n$ square matrices 
whose eigenvalues all have non-positive real part. Note that 
by Lemma~\ref{lem:spectrum},
$\mathbb D(n)$ includes the set $\mathbb G(n)$ of $n\times n$ 
sub-generator matrices (i.e. matrices with non-negative 
off-diagonal elements and non-positive rows).
Recall that $\mathbb C^{n\times m}$ denotes the set 
of $n\times m$ matrices with complex entries. 
Denote by $\mathbb H(n,m)$ the set of $n\times m$ sub-probability matrices (i.e. matrices with non-negative elements and row sums smaller or equal to one). 

Denote by $N$, $N^+$, $\unl N^+$, $N^-$ and $\unl N^-$ 
the number of elements of the sets $E$,
$E^0\cup E^+$, $E^+$, $E^0\cup E^-$, and $E^-$, respectively.
Also, let $\mc H$ denote the
set  
$$
\mc H = \le\{h: E\to \mathbb C\quad: \quad\min_{i\in E}\Re(h(i))\ge 0,\quad
\min_{i\in E^0}\Re(h(i))> 0 \ri\}.
$$

\begin{defin}
Let $h\in\mc H$ and let
$W^+, G^+, W^-$ and $G^-$ be elements of the sets $\mc
\CC^{N\times N^+}$, $\mathbb D(N^+)$, $\CC^{N\times N^-}$ and 
$\mathbb D(N^-)$, respectively. A quadruple $( W^+, G^+, W^-, G^-)$ is
called a {\em matrix Wiener-Hopf factorisation} of $Q_h$ if the following matrix equations are satisfied:
\begin{eqnarray}
\label{systemnoiseWH3} 
\frac{1}{2}\Sigma^2 W^+ (G^+)^2 -
 V W^+  G^+ +   Q_h W^+ &=& O^+,\\
\label{systemnoiseWH3-} 
\frac{1}{2}\Sigma^2 W^- (G^-)^2 +
 V W^-  G^- +   Q_h W^- &=& O^-,
\end{eqnarray}
where 
$O^+$
and
$O^-$
are  zero matrices of size 
$N\times N^+$ 
and
$N\times N^-$.
\end{defin}

\begin{thm}
\label{thm:WienerHopf}
(i) For any $h\in\mc H$,
there exists a unique matrix 
Wiener-Hopf factorisation of $Q_h$, denoted by
$(\eta_h^+,Q_h^+,\eta_h^-, Q_h^-)$.

(ii) If $h=\Re(h)$, then $Q_h^+\in\mathbb G(N^+)$ 
and $Q_h^- \in \mathbb G(N^-)$ are sub-generator matrices, and  
$\eta_h^+\in\mbb H(N,N^+)$ and $\eta_h^-\in\mbb H(N,N^-)$ 
are in block-notation given by
\begin{equation}\label{eq:etahpm}
\eta_h^+ = \begin{pmatrix} I_+ \\ \unl\eta^+_h\end{pmatrix},
\qquad \text{and}\qquad \eta_h^- = 
 \begin{pmatrix} \unl\eta_h^- \\ I_- \end{pmatrix},
\end{equation}
for some matrices $\unl\eta^+_h\in\mbb H(\unl N^-,N^+)$ and 
$\unl\eta_h^-\in\mbb H(\unl N^+,N^-)$, and identity matrices $I_+$
and
$I_-$
of sizes
$N^+\times N^+$
and $N^-\times N^-$.
\end{thm}
{\bf Remarks.} (i) For $h$ given by $h(i)=qI_{E^0}(i)$
with $q>0$ we will also write 
 $(\eta_q^+,Q_q^+,\eta_q^-, Q_q^-)$.
The proof of 
Theorem \ref{thm:WienerHopf} will be given in Section \ref{sec:WH}.

(ii) We allow the vector $h$ to take complex 
values to be able to deal with 
a Laplace inversion using a Browmich integral, 
which involves the integration of the resulting first-passage 
quantities over a curve in the complex plane. Note that, for any 
real-valued $h\in\mc H$, i.e. $h=\Re(h)$, the matrix 
$Q_h$ is a transient generator matrix, since the off-diagonal elements of $Q_h$ are non-negative and $Q_h\1 \neq 0$.

\subsection{First-passage into a half-line}

The marginal distributions of the maximum and minimum as well as the distributions of the first-passage times into a half-line can be described explicitly 
in terms of the matrix Wiener-Hopf factorisation. Denote by $\ovl X_t=\sup_{0\leq s\leq t} X_s$ the running maximum of $X$ at time $t$  and by 
$\unl X_t = \inf_{0\leq s\leq t} X_s$ the corresponding running minimum, and let $T_a^+$ and $T_a^-$ be the first passage times of $X$ into a half-line,
$$
T_a^+ = \inf\{t\ge0: X_t \in (a,\infty)\},\qquad T_a^- = \inf\{t\ge0: X_t \in (-\infty,a)\}.
$$
The distributions of those random variables are related via
\begin{eqnarray*}
\G^+(t) &=& \PP_{x,i}(T_a^+ < t) = \PP_{x,i}(\ovl X_t > a),\\
\G^-(t) &=& \PP_{x,i}(T_a^- < t) = \PP_{x,i}(-\unl X_t > a).
\end{eqnarray*}
If $\mathbf e_q$ denotes a random time that is exponentially distributed with 
parameter $q>0$ and that is 
independent of $(X,Z)$, then the running maximum 
and minimum at time $\mathbf e_q$ follows a phase-type distribution, 
with parameters explicitly given in terms of the matrix Wiener-Hopf factorisation 
$(\eta^+_q, Q_q^+,\eta^-_q, Q^-_q)$.

For any $h\in\mc H$, define 
the $N\times N^+$ and $N\times N^-$ matrices 
$\Phi^+_{h,a}(x)$ and $\Phi^-_{h,a}(x)$ by
\begin{eqnarray}\label{phipm}
\Phi^+_{h,a}(x,i,j) &=& \le[\eta_h^+\exp\left((a - x) Q^+_h\right)\ri](i,j)I_{(-\infty,a]}(x) + \delta_{ij}I_{(a,\infty)}(x), \qquad i\in E, j\in E^+\cup E^0, \\ 
\Phi^-_{h,a}(x,i,j) &=& \le[\eta_h^-\exp\left((x-a) Q^-_h\right)\ri](i,j)
I_{[a,\infty)}(x) + \delta_{ij}I_{(-\infty,a)}(x), \qquad i\in E, j\in E^-\cup E^0,
\label{phimp}
\end{eqnarray}
where $\delta_{ij}$ is a Kronecker delta 
(i.e. $\delta_{ij} = I_{\{i\}}(j)$). 
For $h$ given by $h(i)=qI_{E^0}(i)$
with $q>0$, we will also denote $\Phi^\pm_{h,a}(x)$ by $\Phi^\pm_{q,a}(x)$.

\begin{prop}\label{prop:fipa}
For $q>0$ it holds  that under $\PP_{x,i}$
\begin{eqnarray*}
\ovl X_{\mathbf e_q} - x &\sim& PH(\eta^+_q(i), Q^+_q),\qquad x\in\mbb R, i\in E^0,\\
-\unl X_{\mathbf e_q} + x &\sim& PH(\eta^-_q(i), Q^-_q), \qquad x\in\mbb R, i\in E^0,
\end{eqnarray*}
where $\eta_q^+(i)$ and $\eta_q^-(i)$ are the $i$-th rows of $\eta_q^+$
and $\eta_q^-$. In particular
the Laplace transforms
$\WH{\G}^\pm(q):=\int_0^\infty\te{-qt} \G^\pm(t)\td t$,
for
$q>0$,
are given by the formulae
\begin{eqnarray*}
\WH{\G}^+(q) &=& \frac{1}{q} \sum_{j\in E^0}\EE_{x,i}[\te{-q T_a^+}I_{\{Z_{T_a^+}=j\}}] 
= \frac{1}{q}\sum_{j\in E^0}
\Phi^+_{q,a}(x,i,j), \qquad x,a\in\mbb R, i\in E^0,\\
\WH{\G}^-(q) &=& \frac{1}{q}\sum_{j\in E^0}\EE_{x,i}[\te{-q T_a^-}I_{\{Z_{T_a^-}=j\}}] = 
\frac{1}{q}\sum_{j\in E^0}
\Phi^-_{q,a}(x,i,j), \qquad x,a\in\mbb R, i\in E^0.
\end{eqnarray*}
\end{prop}
{\bf Remarks.} (i) The proof of Proposition \ref{prop:fipa} will be given in Section \ref{sec:WH}.

(ii) Denote by $\Phi^+_{q,\ub}(x,i)$ and $\Phi^-_{q,\lb}(x,i)$ the $N^+$-dimensional and $N^-$-dimensional row vectors with $j$-th 
elements $\Phi^+_{q,\ub}(x,i,j)$ and $\Phi^-_{q,\lb}(x,i,j)$, respectively.
Then Proposition \ref{prop:fipa} implies that,
under the probability measure 
$\PP_{x,i}$, $x\in\mbb R$, $i\in E^0$,
the processes $M^+=\{M^+_t\}_{t\ge 0}$ and $M^-=\{M_t^-\}_{t\ge 0}$, defined by 
\begin{equation*}
M^+_t = \te{-q\le(t\wedge T_\ub^+\ri)}\Phi^+_{q,\ub}\le(X_{t\wedge T_\ub^+},Z_{t\wedge T_\ub^+}\ri),
\qquad
M^-_t = \te{-q\le(t\wedge T_\lb^-\ri)}\Phi^-_{q,\ell}\le(X_{t\wedge T_\lb^-},Z_{t\wedge T_\lb^-}\ri),
\end{equation*}
are row-vectors of bounded martingales. 
Indeed, the Markov property of $(X,Z)$ implies that
\begin{eqnarray*}
\lefteqn{\EE_{x,i}\le[\te{-q T_\ub^+}I_{\le\{Z_{T_\ub^+}=j\ri\}}\bigg|\mc F_t^{(X,Z)}\ri]}\\ &=& I_{\le\{t< T_\ub^+\ri\}}\te{-qt} \EE_{X_t, Z_t}\le[\te{-q T_\ub^+}I_{\le\{Z_{T_\ub^+}=j\ri\}}\ri] + I_{\{t\ge T_\ub^+\}}\te{-q T_\ub^+}I_{\le\{Z_{T_\ub^+}=j\ri\}}\\
&=& I_{\le\{t< T_\ub^+\ri\}}\te{-qt} \EE_{X_t, Z_t}\le[\te{-q T_\ub^+}I_{\le\{Z_{T_\ub^+}=j\ri\}}\ri] + I_{\{t\ge T_\ub^+\}}\te{-q T_\ub^+}\EE_{X_{T_\ub^+}, Z_{T_\ub^+}}\le[\te{-q T_\ub^+}I_{\le\{Z_{T_\ub^+}=j\ri\}}\ri]\\
&=& \te{-q\le(t\wedge T_\ub^+\ri)}\EE_{X_{t\wedge T_\ub^+}, Z_{t\wedge T_\ub^+}}\le[\te{-q T_\ub^+}I_{\le\{Z_{T_\ub^+}=j\ri\}}\ri] \\
&=& \te{-q\le(t\wedge T_\ub^+\ri)}\Phi^+_{q,\ub}\le(X_{t\wedge T_\ub^+}, Z_{t\wedge T_\ub^+}\ri)e^+_j = M^+_t e^+_j,
\end{eqnarray*}
where $e^+_j$ denotes the $j$-th standard basis vector in $\mbb R^{N^+}$ and
we used that $\EE_{x,i}\le[\te{-q T_\ub^+}I_{\{Z_{T_\ub^+}=j\}}\ri]=e^+_j(i)$ if $x\ge \ub$ 
and $i\in E^0$, 
which directly follows from the definitions 
\eqref{phipm} and .

(iii) Suppose that $Q_0=O$. This corresponds to a 
model in which there are no jumps and no switches between the regimes
(i.e. with probability one the process stays in the starting regime 
and evolves as a Brownian motion with drift).
In this case we can identify the matrix Wiener-Hopf factorisation 
in closed form. 
Note that we have
$N^+=N^-=N$
and hence the matrices 
$G^\pm$
and
$W^\pm$
are of dimension
$N\times N$.
If we take 
$W^\pm$ 
to be equal to the identity matrix 
and
$h(i)=R_D(i)+q$,
the matrix equations~\eqref{systemnoiseWH3}--\eqref{systemnoiseWH3-} 
reduce to
$$
\frac{1}{2}\Sigma^2(G^+)^2 -
 V   G^+  = (\Lambda_D + q I) = 
\frac{1}{2}\Sigma^2  (G^-)^2 + V G^-, 
$$
where $I$ is an $N\times N$ identity matrix
(recall that 
the discount rate matrix
$\Lambda_D$
is diagonal and satisfies 
$\Lambda_D(i,i)=R_D(i)$
for all
$i\in E^0$).
These equations are satisfied by the diagonal matrices
\begin{equation}\label{Q+-}
G^+ = \text{diag}(-\w_i^+, i=1,\ldots, N), \qquad 
G^- = \text{diag}(-\w_i^-, i=1,\ldots, N),
\end{equation}
where 
\begin{equation}\label{w+-}
\w_i^\pm = \mp \frac{\mu_i}{\sigma^2_i} + 
\sqrt{\le(\frac{\mu_i}{\sigma_i^2}\ri)^2 + \frac{2(q+r_i)}{\sigma_i^2}}\qquad
\text{and}\quad
r_i:=R_D(i).
\end{equation}
In particular, we obtain the well known fact that the maximum of a Brownian
motion with drift at an independent exponential time is exponentially distributed:
$$\PP_{x,i}\left(\ovl X_{\mathbf e_{q+r_i}}>a\right) = \te{-\omega_i^+(a-x)}\q\text{ for $x < a$}.$$

\subsection{Joint distribution of the maximum and minimum} 
In the previous section we have shown how the marginal 
distribution of the maximum as well as of the
minimum can be explicitly 
expressed in terms of the matrix Wiener-Hopf factorisation. 
Also the {\em joint} distribution of the running maximum and 
minimum,
$$
\psi_{x,i}(t) = \PP_x(\unl X_t > \lb, \ovl X_t < \ub),
$$
can be explicitly identified in terms of the 
matrix Wiener-Hopf factorisation, 
by considering appropriate linear combinations of 
the functions $\Phi^+$ and $\Phi^-$ 
defined in~\eqref{phipm}.

To formulate the result, introduce the matrices $Z^+\in\CC^{N^-\times N^+}$ 
and $Z^-\in\CC^{N^+\times N^-}$ by
%
\begin{eqnarray*}
 Z^+(i,j) & = &  \left[\eta_h^+ \te{ Q^+_h(\ub-\lb)}\right](i,j),\qquad 
 \qquad  i\in E^0\cup E^-,\quad j\in E^0\cup E^+,\\
 Z^-(i,j) & =  & \left[\eta_h^- \te{ Q^-_h(\ub-\lb)}\right](i,j),
 \qquad \qquad  i\in E^0\cup E^+,\quad j\in E^0\cup E^-,
\end{eqnarray*}
and define, for any $h\in\mc H$ and any $x\in\mbb R$, the $N\times N^+$ and $N\times N^-$ matrices $\Psi^+_{h,\lb,\ub}(x)$ and 
$\Psi^-_{h,\lb,\ub}(x)$ 
\begin{eqnarray}\label{eq:twosidedplus}
\Psi^+_{h,\lb,\ub}(x) &=& \le( \eta_h^+\te{ Q^+_h(\ub-x)} - \eta_h^-\te{ Q^-_h
(x-\lb)} Z^+\ri) \le( I -  Z^- Z^+\ri)^{-1}I_{[\lb,\ub]}(x)\\ &\nn \phantom{=}& +\ \Delta^+ I_{(\ub,\infty)}(x), \\
\label{eq:twosidedmin} \Psi^-_{h,\lb,\ub}(x) &=& 
\le( \eta_h^-\te{ Q^-_h(x-\lb)} -
\eta_h^+\te{ Q^+_h
(\ub-x)} Z^-\ri) \le( I -  Z^+ Z^-\ri)^{-1}
I_{[\lb,\ub]}(x)\\ \nn &\phantom{=}& +\ \Delta^- I_{(-\infty,\lb)}(x),
\end{eqnarray}
where $\Delta^+$ and $\Delta^-$ are $N\times N^+$ and $N\times N^-$ matrices with elements
$$
\Delta^+(i,j) = \delta_{ij}, \qquad \Delta^-(i,k) = \delta_{ik}, \qquad\q\q  i\in E, j\in E^+\cup E^0, k\in E^0\cup E^-.
$$
For $h$ given by $h(i)=qI_{E^0}(i)$
with $q>0$, we will also write $\Psi^\pm_{h,\lb,\ub}(x) = \Psi^\pm_{q,\lb,\ub}(x)$.

\begin{prop}
Let
$i\in E^0$
and
\begin{equation*}
\psi_{x,i}^+(t) = \PP_{x,i}(T_{\lb,\ub} < t, X_{T_{\lb,\ub}} \ge \ub),
\qquad 
\psi_{x,i}^-(t) = \PP_{x,i}(T_{\lb,\ub} < t, X_{T_{\lb,\ub}} \leq \lb),
\end{equation*}
where 
$$
T_{\lb,\ub} := \inf\{t\ge 0: X_t \notin [\lb,\ub]\} = T^-_\lb\wedge T^+_\ub.
$$
For any 
$q>0$ 
the Laplace transforms in $t$ of
$\psi_{x,i}^+$
and
$\psi_{x,i}^-$
are given by 
\begin{equation*}
\WH\psi^+_{x,i}(q) =\frac{1}{q} (\Psi^+_{q,x}\1)(i),
\qquad \WH\psi^-_{x,i}(q) = \frac{1}{q} (\Psi^-_{q,x}\1)(i).
\end{equation*}
The Laplace transform 
of
$\psi_{x,i}(t) = \PP_{x,i}(\unl X_t > \lb, \ovl X_t < \ub)$ is hence of the form
\begin{equation}\label{twoexit}
\WH\psi_{x,i}(q) = \WH\psi^+_{x,i}(q) + \WH\psi^-_{x,i}(q).
\end{equation}
\end{prop}
{\bf Remarks.} (i) If $Q_0=O$, then $\eta_q^\pm$ are identity matrices and 
the following identities hold:
\begin{eqnarray*}
\Psi^+_{q,x} & = & (\te{Q_q^+(\ub-x)} - \te{Q_q^-(x-\lb)}\te{Q^+_q(\ub-\lb)})
(I - \te{Q^-_q(\ub-\lb)}\te{Q^+_q(\ub-\lb)})^{-1},\\
\Psi^-_{q,x} & = & (\te{Q_q^+(\ub-x)} - \te{Q_q^+(x-\lb)}\te{Q^-_q(\ub-\lb)})
(I - \te{Q^+_q(\ub-\lb)}\te{Q^-_q(\ub-\lb)})^{-1},
\end{eqnarray*}
where 
$Q^\pm_q$
are diagonal matrices given in~\eqref{Q+-}.
In particular, we find the well known two-sided exit identity 
for Brownian motion with drift:
$$
\WH\psi_{x,i}(q)= \EE_{x,i}[\te{-q \tau_{\lb,\ub}}I_{\{\tau_\ub < \tau_\lb\}}]
= 
\frac{\te{\w_i^+(x-\lb)} - \te{\w_i^-(x-\lb)}}{\te{\w_i^+(\ub-\lb)} -
\te{\w_i^-(\ub-\lb)}},
$$
where 
$\w^\pm_i$
is given in~\eqref{w+-}
with 
$r_i=0$.

(ii) In the case that no volatility is present ($\Sigma\equiv 0$) 
the identities simplify and we find the expressions in \cite{APU}.

(iii) Under $\PP_{x,i}$, $x\in\mbb R$, $i\in E^0$,
the process $\WT M^+=\{\WT M^+_t\}_{t\ge 0}$ 
defined by 
\begin{equation*}
\WT M^+_t = \te{-q\le(t\wedge T_{\lb,\ub}\ri)}\Psi^+_{q,\lb,\ub}\le(X_{t\wedge T_{\lb,\ub}},Z_{t\wedge T_{\lb,\ub}}\ri),
\end{equation*}
is a row-vector of bounded martingales, where we denoted
by $\Psi^+_{q,\lb,\ub}(x,i)$ 
the $N^+$-dimensional 
row vector with $j$-th 
element $\Psi^+_{q,\lb,\ub}(x,i,j)$.

\begin{pr} Let $q>0$ and define $$g^+(x,i) = \Psi_{q,\lb,\ub}^+(x,i).$$
In view of the fact that $\WT M^+$ is a bounded martingale, it holds that
\begin{eqnarray*}
g^+(x,i) &=& \sum_{j\in E^0\cup E^+}\EE_{x,i}\le[\te{-q\tau}
\Psi^+_{q,\lb,\ub}(X_{\t}, Z_\t,j)I_{\{\tau<\infty\}}\ri] \\
&=& \sum_{j\in E^0\cup E^+}\EE_{x,i}\le[\te{-q\tau}I_{\{X_\t \ge \ub, Z_\t = j\}}\ri]\\
&=& q \int_0^\infty \te{-q t}\PP_{x,i}(\tau < t, X_\t \ge \ub)\td t 
 = q \WH\psi^+_{x,i}(q).\end{eqnarray*}
Here we used that, in view of the definitions \eqref{eq:twosidedplus} 
and \eqref{eq:etahpm} 
of $\Psi^+_{q,\lb,\ub}$ and $\eta_q^+$, the function $g^+$ satisfies: 
\begin{eqnarray*}
g^+(x,i) &=& 1 \qquad \text{if $x\ge\ub$, $i\in E^0$}\\
g^+(x,i) &=& 0 \qquad \text{if $x\leq\lb$, $i\in E^0$.}
\end{eqnarray*}
The expression for $\WH\psi_{x,i}^-$ can be derived by a similar reasoning.
\end{pr}

\subsection{Valuing a double-barrier rebate option}
A double-barrier rebate option pays a constant rebate $L$ at the moment 
$\tau$ one of the barrier levels is crossed, if this happens before maturity $T$. By standard arbitrage pricing arguments the Laplace transform $\WH v_{reb}$ in maturity of the price of $v_{reb}$ such an option is given by 
\begin{equation}\label{eq:reb}
\WH v_{reb}(q) = \frac{1}{q}\EE_{x,i}\left[\le(B^D_\tau\ri)^{-1}
\exp\left(- q\tau \right)\ri]
\end{equation}
We will find below the following more general quantity, 
which will also be employed in the sequel:
\begin{eqnarray}
\label{eq:TheCoordinates}
H^{x,i}(q,u,j) & := & \EE_{x,i}\left[\le(B^D_\tau\ri)^{-1}\exp\left(\cu uX_\tau- q\tau \right) I_{\{Z_\tau=j\}}\right]
\quad\text{for}\quad j=1,\ldots,N.
\end{eqnarray}
In the following result an explicit expression is given for the 
quantity $H_j^{x,i}$ in terms of the matrix Wiener-Hopf factorisation, which is an extension of the identity \eqref{twoexit}.

%
We denote by $E_i^+$ and $E_i^-$, for $i\in E^0=\{1, \ldots, N\}$, 
the parts of the state-space $E$ corresponding to the 
blocks $B_i^+$ and $B_i^-$ in the matrix $Q_0$ in \eqref{eq:Q0}.
Recall that the definition of $\alpha_i^\pm$ was given in \eqref{eq:SmallestEigVal}.

\begin{thm}
\label{thm:MainEmbed}
For any $h\in\mathcal H$, $i,j\in E^0$, and $u\in\CC$
that satisfies $\Im(u)\in(-\alpha_j^+,\alpha_j^-)$ 
it holds that
\begin{equation}\label{eq:H}
H^{x,i}_j(h,u) = \le(\Psi^+_{h,x}k^+_{j,u}\ri)(i) + \le(\Psi^-_{h,x}k^-_{j,u}\ri)(i),
\end{equation}
where the column vectors $k^+_{j,u}=(k_{j,u}^+(i), 
i\in E^+\cup E^0)$ and 
$k^-_{j,u}=(k_{j,u}^-(i), i\in E^0\cup E^-)$ are given by
\begin{eqnarray*}
k^+_{j,u}(i) &=& \te{u\ub} \cdot
\le\{\begin{array}{ll}
1,\q & \text{if $i=j\in E^0$},\\
\le((-u I_j^+ - B^+_j)^{-1} (-B_j^+)\1\ri)(i),\q & 
\text{if $i\in E^+_j$},\\ 
0,\q&\text{otherwise}\\
\end{array}\ri.
\\
k^-_{j,u}(i) &=& \te{u\lb}\cdot
\le\{\begin{array}{ll}
1,\q & \text{if $i=j\in E^0$},\\
\le((u I^-_j - B^-_j)^{-1} (-B^-_j)\1\ri)(i),\q& 
\text{if $i\in E^-_j$},\\
0,\q & \text{otherwise}\\ 
\end{array}\ri.
\end{eqnarray*}
where 
$ I^+_j$ and $I^-_j$ are $|E_j^+|\times |E_j^+|$ and $|E_j^-|\times |E_j^-|$ identity matrices.
\end{thm}
 The proof is given in Section \ref{sec:WH}.

\section{Double-no-touch and other barrier options}
\label{sec:DNT_KOC}
In this section we show how the prices of  double-no-touch
and double knock-out call options can be expressed in terms 
characteristic function $H$ of the process at the corresponding 
first passage time, which was identified in  
Theorem~\ref{thm:MainEmbed}. 

A \textit{double-no-touch}
is a derivative security 
that pays one unit of the underlying asset
at expiry 
$T$ 
if the
underlying 
asset price
does not leave the interval 
$[\exp(\lb),\exp(\ub)]$
during the time period
$[0,T]$, where $\lb<\ub$.
Similarly a \textit{double knock-out call option}
struck at 
$K$
delivers the payoff
$(S_T-K)^+:=\max\{S_T-K,0\}$
if throughout the life of the option 
the asset price stays within the interval
$[\exp(\lb),\exp(\ub)]$.
%
The arbitrage free prices 
for a double-no-touch 
and 
a double knock-out call option 
are respectively  given by
\begin{eqnarray}
\label{eq:DNT_Price}
D_{x,i}(T) & =  & \EE_{x,i}\left[\frac{I_{\{\tau>T\}}}{B_T^D} \right],\\
C_{x,i}(k,T) & =  & \EE_{x,i}\left[\frac{I_{\{\tau>T\}}}{B_T^D}(S_T-K)^+\right],
\label{eq:DKOC_Price}
\end{eqnarray}
%
where $k=\log K$ is the log-strike and $\tau$ is the first time the process
$S$ leaves the interval $[\exp(\lb),\exp(\ub)]$
or equivalently
\begin{eqnarray}
\label{eq:FirstPassageTime}
\tau := \inf\{t\ge0: X_t\notin[\lb,\ub]\}.
\end{eqnarray}
We will find it more convenient to consider the double-touch-in 
and the knock-in call whose values $v_{dti}$ and $v_{kic}$, 
as function of maturity $T$ and log strike $k=\log K$,
are given by
\begin{eqnarray}
\label{eq:DNT_forLaplace}
v_{dti}(T) &:=& \EE_{x,i}\left[\frac{I_{\{\tau\leq T\}}}{B_T^D} \right]
=  \EE_i\left[(B_T^D)^{-1}\right] - D_{x,i}(T),\\
%
v_{kic}(T,k) &:=& \EE_{x,i}\left[\frac{I_{\{\tau\leq T\}}}{B_T^D}(S_T-K)^+\right]
 = 
\EE_{x,i}\left[\le(B_T^D\ri)^{-1}(S_T-\te{k})^+\right] - C_{x,i}(k,T).
\label{eq:KOC_forLaplace}
\end{eqnarray}
Since the zero-coupon bond price as well as the call option price 
have already been identified, the problem of calculating the double-no-touch and knock-out call prices thus reduces to identifying $v_{dti}$ and $v_{kic}$. In general, no closed form expressions are known for these function in terms of elementary functions. Below we show that the Laplace transform in $T$ of $v_{dti}(T)$ as well as the joint Fourier-Laplace transform in $(k,T)$ of $v_{kic}$ can be identified explicitly in terms of the parameters that define the log-price process $X$.
Both transforms will be identified in terms of 
Laplace transform $\WH F_{x,i}(u,q)$ in $T$ of the function 
$T\mapsto F_{x,i}(u,T)$ given by
\begin{eqnarray}
\label{eq:FirstPassageCharF}
F_{x,i}(u,T):= \EE_{x,i}\left[\frac{I_{\{\tau\leq T\}}}{B_T^D}\exp(\cu uX_T)\right].
\end{eqnarray}
The Laplace transform $\WH F_{x,i}(u,q)$ in turn will be given in terms of
the $N_0$-vector $H^{x,i}(q,u)$ whose coordinates are given by
\begin{eqnarray*}
H^{x,i}_j(q,u) & := & \EE_{x,i}\left[\exp\left(\cu uX_\tau-\int_0^\tau(R_D(Z_s)+q)ds\right) I_{\{Z_\tau=j\}}\right]
\quad\text{for}\quad j=1,\ldots,N_0.
\end{eqnarray*}

\begin{thm}
\label{thm:Projection}
For any $q>0$ and $\xi$ with $\Im(\xi)<0$ it holds that
\begin{eqnarray}
\label{dti}
\WH v_{dti}(q) &=& \WH F_{x,i}(0,q),\\
\WH v_{kic}^*(q,\xi) &=& \frac{1}{\cu\xi - \xi^2}\WH F_{x,i}(\xi-\cu,q).
\end{eqnarray}
Here, $\WH F_{x,i}(u,q)$ is given by
\begin{eqnarray}
\label{eq:FinalLapTrnsfOfF}
\WH F_{x,i}(u,q) & = & \left(H^{x,i}(q,u)\right)'(qI + \Lambda_D-K(u))^{-1}\1,
\end{eqnarray}
for all $q\in\CC$,
such that
$\Re(q)> q^* = \max\{\Re(\psi_i(u))-R_D(i)\>:\>i=1,\ldots,N_0\}$.
\end{thm}
{\bf Remarks.} (i) Note that if
$u=0$, then $q^*\leq 0$, as the interest rates $R_D(i)$ is assumed to be non-negative.

(ii) The Laplace transform in \eqref{eq:FinalLapTrnsfOfF} 
can be inverted by evaluating the Bromwich integral
\begin{equation}
F_{x,i}(u,t)  = \frac{1}{2\pi} \int_{c-\cu\infty}^{c+\cu\infty}\te{tq}
\left(H^{x,i}(q,u)\right)'(qI + \Lambda_D-K(u))^{-1}\1
\td q,
\end{equation}
for any $c>q^*$. An efficient algorithm to approximate this integral can e.g. be found in Abate and Whitt \cite{AW}.

\begin{pr} First note that by a similar calculation to the one in the proof
of Proposition~\ref{Prop:FourTrnsformVanilla}
we find that the Fourier-transform in log-strike $k$ of $\te{\alpha k}v_{kic}(T,k)$
can be expressed in terms of $F$ by
$$
\int_{\mathbb R}\te{(\cu v + \a)k}v_{kic}(T,k)\td k = 
\frac{1}{(\a+\cu v)(1+\a+\cu v)} \mathbb E_{x,i}\le[\frac{I_{\{\tau\leq T\}}}{B_T^D} \exp\{(1 + \a + \cu v)X_T\}\ri],
$$
for any $\alpha>0$.

Assume next that 
$q>0$ is real with $q>q^*$.
The Laplace transform $\WH F_{x,i}(u,q)$
has the following well-known equivalent probabilistic representation 
\begin{eqnarray}
\WH F_{x,i}(u,q) 
& = & \EE[F_{x,i}(u,e_q)]/q,
\label{eq:ExpTimeLT}
\end{eqnarray}
where
$e_q$
is an exponential random variable with parameter
$q$
which is independent of the process
$(X,Z)$.

By conditioning on 
$\FF_\tau$
and applying 
strong Markov property 
at 
$\tau$
of the process
$(X,Z)$
together with
Lemma~\ref{lem:ExpGen}
we find that
\begin{eqnarray}
\EE[F_{x,i}(u,e_q)] & = & 
\EE_{x,i}\left[\frac{\exp(\cu u X_\tau)}{B_\tau^D}
\EE\left[I_{\{\tau\leq e_q\}}\exp\left(-\int_\tau^{e_q}R_D(Z_s)\td s+\cu u(X_{e_q}-X_\tau)\right) \Big\lvert\FF_\tau\right]\right]\nonumber\\
& = & \EE_{x,i}\left[\frac{\exp(\cu u X_\tau)}{B_\tau^D}
h(Z_\tau,q,u)e^{-q\tau}\right]\nn\\
&=& \sum_j H_j^{x,i}(q,u) h(j,q,u),
\label{eq:LaplaceTransBar}
\end{eqnarray}
where the value
$h(j,q,u)$
for each state 
$j\in E^0$
of the Markov chain 
$Z$
is given by
\begin{eqnarray}
\label{eq:def_h}
h(j,q,u) & := & \EE_{0,j}\left[\exp\left(-\int_0^{e'_q}R_D(Z_s)\td s+\cu uX_{e'_q}\right)\right]\\
&=& \le[ (qI + \Lambda_D-K(u))^{-1}\1 \ri](j)\nn
\end{eqnarray}
for some exponential random variable 
$e'_q$
with parameter
$q$
that is independent of the Markov process
$(X,Z)$, where the second equality follows from
Lemma~\ref{lem:h}
and Theorem~\ref{thm:CharFun}.

A key observation that follows from the
representation~\eqref{eq:LaplaceTransBar}
is that the function
$$q\mapsto \EE[F_{x,i}(u,e_q)]/q$$
has a holomorphic extension to the complex half-plane
$\{q\in\CC\>:\>\Re(q)>q^*\}$
which therefore\footnote{If two holomorphic 
functions defined on a connected open set 
$\Omega$
in 
$\CC$
coincide on a subset with at least one accumulation point in
$\Omega$,
then they coincide on the entire 
$\Omega$.
For a proof of this well-known statement
see~\cite{Rudin}, page~208,
Theorem~10.18.}
coincides on this domain with the Laplace transform. 
Thus, Equation \eqref{eq:LaplaceTransBar} 
holds for $q$ in this domain, and the proof is complete.
\end{pr}

\section{Embedding of the process $(X,Z)$}
\label{sec:FE}

%

In this section and the next we will provide a proof of the 
matrix Wiener-Hopf factorisation and its corollaries 
derived in previous sections. We will proceed in two steps: 
\begin{itemize}
\item[(i)] Reduction of the first-passage problems of $X$ 
over constant levels, 
which will involve overshoots and undershoots due to the jumps of $X$, 
to the first-hitting problem of a constant level by a 
regime-switching Brownian motion, employing a classical argument 
for the embedding of phase-type jumps 
(see Asmussen \cite{Asmussenruin}), which we review in this section.

\item[(ii)] Solution of the first-passage problem of regime-switching Brownian motion
via a characterisation of the dynamics of the ladder processes, which is carried out in Section \ref{sec:WH}.
\end{itemize}

%
%

Let $Y$ be a continuous time 
Markov chain with finite
state space $E\cup \partial$ and generator restricted to $E$ given by 
$Q_0$, 
where $\partial$ is an absorbing
cemetery state and $E$ and $Q_0$ will be specified shortly, 
and denote by 
$\xi = \{\xi_t, t\ge 0\}$ the Markov modulated Brownian motion given by
\begin{equation}\label{eq:fluidem}
\xi_t = x + \I_0^ts(Y_s)\td W_s + \I_0^t m(Y_s)\td s,
\end{equation}
where $x\in\RR$ is the starting point of $\xi$ and  $s$ and $m$ are functions from $E\cup \partial$ to $\RR$ also to be specified shortly.
The couple $(\xi,Y)$ defined as such is a 
two-dimensional strong Markov process. In the sequel we will denote by 
$\WT{\PP}_{x,i}$ and 
$\WT{\EE}_{x,i}$  the conditional probability 
$\WT{\PP}_{x,i} = \WT{\PP}[\cdot|\xi_0=x,Y_0=i]$ and 
conditional expectation $\WT{\EE}_{x,i} = \WT{\EE}[\cdot|\xi_0=x,Y_0=i]$ respectively.  

\begin{figure}[t]
\centering
\input{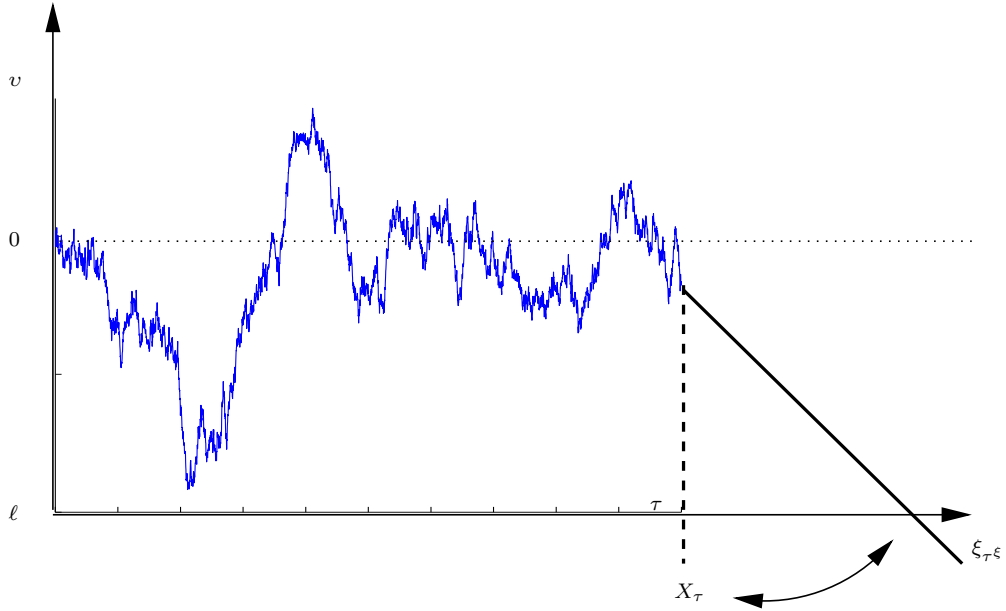}
\caption{Shown is a sample path of $X$ until the first time $\tau$
that $X$ exits the interval $[\lb,\ub]$. The process $\xi$ has 
no positive jumps
and always hits a level at first-passage.} \label{fig:embed}
\end{figure}

Let the state-space
$E$
be as in Section~\ref{sec:AD}.
In other words
$E=E^-\cup E^0\cup E^+$
where
$E^0$
is the state-space of the chain
$Z$
and
$E^+$
and
$E^-$
are given in~\eqref{eq:E}.
The generator 
$Q_0$
is given in~\eqref{eq:Q0}
and 
$m(i):=\Lambda_V(i,i)$,
$s(i):=\Lambda_\Sigma(i,i)$
for all
$i\in E$,
where  the matrices
$\Lambda_V$,
$\Lambda_\Sigma$
are defined in~\eqref{eq:SiV}.
Thus, while the chain $Y$ is in state $j\in E^0$, $\xi$ evolves 
as a Brownian motion with drift $m(j)$ and 
volatility $s(j)$, and while $Y$ takes values in 
$E^+$ and $E^-$, the path of $\xi$ is linear with slope $+1$ or $-1$.
Informally, a path of 
$X$ can be obtained from a path of $\xi$ by replacing these 
stretches of unit slope by jumps of the same length, 
as is illustrated in Figure \ref{fig:embed}.
These linear increasing and decreasing stretches of path of $\xi$ 
thus correspond to the positive and negative jumps 
of $X$, respectively.  


More formally, by appropriately time-changing $(\xi,Y)$ 
a stochastic process can be constructed that has the same law as $(X,Z)$.
Denote by
$$T_0(t)=\I_0^tI_{\le\{Y_s\in E^0\ri\}}\td s\qquad\text{and}
\qquad T_0^{-1}(u) = \inf\{t\ge0: T_0(t)>u\},
$$
the time before $t$ spent by the chain 
$Y$ 
in 
$E^0$ 
and its right-continuous
inverse respectively. 
It is clear from the definition of the generator
$Q_0$
in~\eqref{eq:Q0}
that when the chain
$Y$
jumps from any of the states in
$E^+\cup E^-$
to a state 
$j\in E^0$,
it must have been in the state
$j$
just before it had left
$E^0$.
Therefore 
the form of the matrix~\eqref{eq:Q0}
implies that
the process
$Y\circ T_0^{-1}$,
which simply ignores all 
excursions of
$Y$
into
$E^+\cup E^-$,
is a Markov chain with the same generator 
as 
$Z$.
Furthermore 
it is 
straightforward to verify that 
$\xi \circ T_0^{-1}$ 
is a regime-switching jump-diffusion 
and 
the law of the process $(\xi \circ T_0^{-1}, Y\circ T_0^{-1})$ under 
$\WT{\PP}_{x,i}$ 
is equal to that of $(X,Z)$ under $\PP_{x,i}$, for all $x\in\RR$ and $i\in E^0$. 

If we define
the first-passage time
\begin{equation}\label{eq:tauxi}
\tau^\xi = \inf\{t\ge 0: \xi_t\notin[\lb,\ub] 
\quad\text{and}\quad Y_t\in E^0 \}  
\end{equation}
it follows that random variables 
$T_0(\tau^\xi)$
and the stopping time 
$\tau$
defined in~\eqref{eq:FirstPassageTime}
have the same distribution.
In particular, with the extension of $h$ to $E$ that puts $h(i)=0$ 
for $i\notin E^0$ and that we will also denote by $h$,
it holds that
$$
\le(\xi_{\tau^\xi}, \int_0^{\tau^\xi} h(Y_s)\td s, Y_{\tau^\xi}\ri)
\ \text{under $\WT{\PP}_{x,i}$ has the same distribution  as}\
\le(X_{\tau}, \int_0^{\tau}h(Z_s)\td s, Z_{\tau}\ri)
\text{under $\PP_{x,i}$}
$$
for $x\in\RR$ and $i\in E^0$.
The function $H^{x,i}_j$ defined in \eqref{eq:TheCoordinates} can 
thus be expressed in terms of the embedding
$(\xi,Y)$ as follows:
\begin{equation}\label{eq:Hxij}
H^{x,i}_j(h,u)  =  
\WT{\EE}_{x,i}\left[\exp\left(u\xi_{\tau^\xi} - \int_0^{\tau^{\xi}} h(Y_s)
\td s\right) I_{\{Y_{\tau^\xi}=j\}}\right]
\quad\quad
i,j\in\ E^0.
\end{equation}

\section{Ladder processes}
\label{sec:WH}
This section is devoted to the proof of 
existence and uniqueness of the $h$-matrix Wiener-Hopf 
factorisation, for any $h\in\mc H$. For real 
$h\in\mathcal H$ the vector $h$ has the probabilistic 
interpretation of a vector of state-dependent rates of discounting, and 
in this case the matrix Wiener-Hopf factorisation has a probabilistic 
interpretation in terms of ladder processes. The existence and uniqueness results are extended to the case of general complex $h\in\mc H$ 
by analytical continuation arguments. It is of interest to consider 
the case of complex entries, as the Laplace transforms 
of barrier option prices are expressed in terms of the 
matrix Wiener-Hopf factorisation as we saw above
and some widely used Laplace transform inversion algorithms are based on the Bromwich (complex) integral representation.
 
A classical probabilistic approach to characterisation of the 
joint distribution of $\tau_a^+$ and the position of $Y$ 
at $\tau_a^+$ is to consider the up-crossing 
ladder process of $(\xi,Y)$, 
defined as follows:

\begin{defin} The {\em up-crossing ladder process} $Y^+=\{Y^+_t\}_{t\ge 0}$ of $(\xi,Y)$ is given by
\begin{equation}
 \label{eq:jtilde}  Y^+_a := 
\begin{cases}
 Y\le(\t^+_a\ri), &\text{if $\t_a^+<\infty$},\\
\partial, &  \text{otherwise},
\end{cases}
\end{equation}
where $\partial$ is a graveyard state and
$$\t^+_a = \inf\{s\ge0: \xi_s  > a\}
$$
with $\inf\emptyset=\infty$.
\end{defin}

\no{\bf Remark.}  In the case that the original chain is killed at (state-dependent, 
real-valued) 
rate $h$, the up-crossing ladder process can be defined as follows. 
Recall from Markov chain theory that 
the chain $Y^h$ with state-space 
$E\cup\{\partial\}$ and generator matrix 
$$
\begin{pmatrix}
Q_h & -\Lambda_h\1\\
\mbf 0 & 0
\end{pmatrix} \qquad \text{where $Q_h:=Q_0 - \Lambda_h$ and 
$\Lambda_h$ is the diagonal matrix with $(\Lambda_h)_{ii}=h_i$}
$$
has the same distribution as the chain $Y$ killed (i.e. sent to the graveyard state $\partial$) independently at rate $h(i)$ when $Y_t=i$. In particular, for $i,j\in E$ and $t\ge 0$ it holds that
\begin{equation}\label{eq:semiid}
 \EEW_i\le[\te{-\int_0^th(Y_s)\td s}I_{\{Y_t=j\}}\ri] = \WT{\PP}(Y^h_t=j|Y^h_0=i) = \exp(Q_ht)(i,j).
\end{equation}
If we denote by 
$\xi^h=\{\xi^h_t\}_{t\ge 0}$ the process defined by \eqref{eq:fluidem} 
driven by the ``killed'' chain $Y^h$ instead of $Y$, then 
the up-crossing ladder process $Y^{h+}$ is given by
$$
Y^{h+}_t = \le\{\begin{array}{cc} Y^h_{\tau_t^{h+}} & \text{if $\tau^{h+}_t<\infty$},\\
\partial & \text{otherwise},
\end{array}\ri.
$$
where $\tau^{h+}_a$ is the stopping time defined by 
$$
\t^{h+}_a = \inf\{s\ge0: \xi^h_s  > a\}
$$
with
$\inf\emptyset = \infty$.

\begin{prop} 
\label{eq:PropMarkovPropertyY_Plus}
The process 
$ Y^+$ is a Markov chain with state-space
$E^0\cup E^+\cup\{\partial\}$.
\end{prop}

\no{\bf Remarks.} (i) We denote by $Q_0^+$ the generator of $Y^+$ restricted to
$E^0\cup E^+$ and by $\eta^+_0$
the initial distribution,  given by 
\begin{eqnarray}
\label{etaijpm} \eta^+_0(i,j) &=& \WT\PP_{0,i}\le[ Y_0^+ = j,
\t_0^+<\infty\ri]\qquad \text{for $i\in E$, $j\in E^+\cup E^0$}
\end{eqnarray}

(ii) More generally, for any real valued $h\in\mc H$ it holds that $Y^{h+}$ is a Markov chain, with generator, restricted 
to $E^0\cup E^+$ denoted by $Q_h^{+}$
and initial (sub-)probability distribution by 
$\eta_h^+=(\eta_h^+(i,j), i\in E, j\in E^+\cup E^0)$ 
with
\begin{eqnarray}
\label{eq:etah+} \eta^+_h(i,j) &=& \EEW_{0,i}\le[
\te{-\int_0^{\t_0^+}h(Y_s)\td s}I_{\{ Y_0^+ = j,
\t_0^+<\infty\}}\ri]\quad \text{for $i\in E$, $j\in E^+\cup E^0$}.
\end{eqnarray}
We have the following identities
analogous to \eqref{eq:semiid}:
\begin{eqnarray}\nn
 \EEW\le[\te{-\int_{0}^{\tau_t^+}h(Y_s)\td s}I_{\{Y_{\tau_t^+}=j, \tau^+_t<\infty\}}\bigg| Y_{0}=i\ri] &=&  \EEW\le[\te{-\int_{\tau^+_0}^{\tau_t^+}h(Y_s)\td s}I_{\{Y_{\tau_t^+}=j, \tau^+_t<\infty\}}\bigg|  Y_{\tau^+_0}=i\ri]\\
\nn &=&
\WT{\PP}\le(Y^{h}_{\tau_t^{h+}}=j, \tau_t^{h+}<\infty\big|Y^{h}_{\tau_0^{h+}}=i\ri)\\
\nn &=&\WT{\PP}(Y^{h+}_t=j|Y^{h+}_0=i)\\
& = & \exp(Q_{h}^+t)(i,j),
\qquad i,j\in E^0\cup E^+, \label{eq:semiid+}
\end{eqnarray}
where we wrote  $\WT{\PP}[\cdot|A] :=\WT{\PP}[\cdot| A\cap\{\xi_0=0\}]$
and $\WT{\EE}[\cdot|A] :=\WT{\EE}[\cdot| A\cap\{\xi_0=0\}]$ to  simplify the notation.
Note that the first equality holds since 
$\tau^+_0=0$
if the chain 
$Y$
starts at
$i\in E^0\cup E^+$.
In particular, 
From \eqref{eq:semiid+} we find that for $i,j\in E^0\cup E^+$
\begin{eqnarray}
\label{eq:Qh+ij}
Q^{+}_h(i,j) &=& \lim_{t\downarrow 0} \frac{1}{t} \EEW_{0,i}\le[\te{-\int_0^{\tau_t^+}h(Y_s)\td s}I_{\{Y_{\tau^+_t}=j, \tau^+_t<\infty\}}\ri], \qquad i\neq j,\\
\label{eq:Qh+ii}
Q^{+}_h(i,i) &=& \lim_{t\downarrow 0} \frac{1}{t} \le\{\EEW_{0,i}\le[\te{-\int_0^{\tau_t^+}h(Y_s)\td s}I_{\{Y_{\tau^+_t}=i, \tau^+_t<\infty\}}\ri] - 1\ri\}.
\end{eqnarray}

\begin{pr}
Let $f$ be any bounded real-valued Borel function and let $b<a.$
Denote by $\mc F_a=\sigma\{Y_u^+\}_{u\leq a}$ and $\mc G_t=\sigma\{Y_s\}_{s\leq t}$ the sigma algebras generated by $Y^+_u$ up to time $a$ and by $Y_s$ 
up to time $t$. Observing that $\mc F_b\subset \mc G_{\tau_b}$ and using 
the spatial homogeneity and continuity of $\xi$ 
and the strong Markov property of $(\xi,Y)$ we find that
\begin{eqnarray*}
\EEW_{x,i}[f(Y^+_a)|\mc F_b] &=& \EEW_{x,i}\le[\EEW_{x,i}\le[f(Y_{\tau^+_a})|\mc G_{\tau_b}\ri]\,\big|\mc F_b\ri]\\
&=& \EEW_{x,i}\le[\EEW_{b,Y_{\tau^+_b}}[f(Y_{\tau^+_{a}})]\,\big|\mc F_b\ri]\\
&=& \EEW_{x,i}\le[\EEW_{Y^+_{b}}[f(Y^+_{a-b})]\,\big|\mc F_b\ri] = \EEW_{Y^+_{b}}[f(Y^+_{a-b})]
\end{eqnarray*}
where we wrote $\EEW_{j}=\EEW_{0,j}$.
\end{pr}

For any $h\in\mathcal H$,
we define the square-matrix (resolvent) functions
$u\mapsto R^+_h(u)$ 
of dimension
$N^+$
on the complex half-plane
$\CC_{>0}$
by
\begin{eqnarray}
\label{eq:R+} 
R^+_h(u)(i,j) &=& \EEW_{0,i}\le[\int_0^\infty
\te{-uy-\int_0^{\t_y^+}h(Y_s)\td s}I_{\{ Y_y^+ = j,
\t_y^+<\infty\}}\td y\ri]\quad \text{for $i,j\in E^+\cup E^0$ and $u\in\CC_{> 0}$}.
\end{eqnarray}

The matrix $Q_h^{+}$ can then be defined for any $h\in\mc H$, as follows:

\begin{lem}\label{lem:complex} 
Let $h\in\mathcal H$ and $a\ge 0$ and
define the matrix 
$Q_h^+\in\mathbb D(N^+)$ 
by
\begin{eqnarray}
\label{eq:Qhij+}
Q_h^+(i,j) &=& -\omega_i^+ I_{\{i=j\}}+
\begin{cases}
\frac{2}{\sigma_i^2}(\widetilde{Q}\eta_h^+ R_h^+(\omega_i^-))(i,j) & \text{if $i\in E^0$},\\
(\widetilde{Q}\eta_h^+)(i,j) & \text{if $i\in E^+$},\\
\end{cases}
\end{eqnarray}
where $q_i = -Q_0(i,i)$,
$h_i = h(i)$, 
$\widetilde{Q}=Q_0+\mathrm{diag}\{q_i\>:\> i \in E\}$
and
\begin{equation}
\omega_i^+ = 
\begin{cases}
\displaystyle 
F\le(\frac{ \mu_i}{\sigma^2_i},\frac{q_i + h_i }{\sigma^2_i}\ri)
& \text{if $i\in E^0$} \\
\displaystyle q_i + h_i & \text{if $i\in E^\pm$}
\end{cases}
\end{equation}
where
\begin{equation}
F(\nu,\theta) := - \nu + \sqrt{\nu^2 + 2\theta}\qquad \nu\in\mbb R,\, \theta\in\mbb C_{> 0}.
\end{equation}
Then it holds that
\begin{eqnarray}
\label{eq:Qh+}
\EEW_{0,i}\le[
\te{-\int_0^{\t_a^+}h(Y_s)ds}I_{\{ Y^+_a=j,\t_a^+<\infty\}}\ri] 
&=& 
\exp(Q_h^+a)(i,j)\qquad i,j\in E^0\cup E^+.
\end{eqnarray}
\end{lem}
{\bf Remarks.} (i) Note that Proposition~\ref{prop:fipa}
follows as a direct consequence of  
Equation \eqref{eq:Qh+}
since
$$\PP_{x,i}\left(\ovl X_{\mathbf{e}_q}>a\right) = \PP_{x,i}\left(Y^{q+}_\zeta>a\right),$$
where
$\zeta$
is the life time of the killed up-crossing ladder process.

(ii) Equation \eqref{eq:Qh+}  
yields in particular the joint Laplace transform 
of the vector 
\begin{equation}
\mbf Z^+_a = \le(\int_0^{\tau_a^+} I_{\{Y_s=i\}}\td s,\ i\in E\ri),
\end{equation}
whose components record the length of time spent by $Y$ in 
each of the states in $E$, until the moment $\tau^+_a$ of first-passage.

(iii) The \textit{down-crossing ladder process} $Y^-$, defined as 
the up-crossing ladder process of $(-\xi,Y)$, is a Markov chain 
with generator restricted to $E_0\cup E^-$ denoted by $Q_0^-$. 
Analogously to 
equation~\eqref{eq:Qhij+},
for any $h\in\mc H$ a matrix $Q_h^-$ can be defined, 
which satisfies
\begin{equation}
\label{eq:Qh-}
\EEW_{0,i}\le[
\te{-\int_0^{\t_a^-}h(Y_s)ds}I_{\{ Y^-_a=j, \t_a^-<\infty\}}\ri] =
\exp(Q_h^-a)(i,j)\qquad i,j\in E^0\cup E^-.
\end{equation}


\subsection{Proof of Lemma~\ref{lem:complex}}
\label{Subsec:Proof_Lem}
%

We will show that the limits in \eqref{eq:Qh+ij}--\eqref{eq:Qh+ii} 
in fact exist for any $h\in\mathcal H$, and identify these. 
Let 
$$\rho=\inf\{t\ge0: Y_t\neq Y_0\}$$ 
be the first time that the chain
$Y$ 
jumps and 
$\tau_a^i=\inf\{t\ge 0: X^i_t>a\}$ 
the first-passage time of 
$X^i_t:=\mu_i t + \sigma_i W_t$ 
(recall that if 
$i\in E^\pm$,
then
$\mu_i=\pm1$
and
$\sigma_i=0$)
over the level $a$ and let 
$\mathbf e_i$ 
be an exponential random time with mean $1/q_i$ (with $q_i=-Q_0(i,i)$) that is independent of $X^i$. 
In view of the definition of $\xi$
(i.e. the first jump time 
$\rho$
of
$Y$
is independent of 
$X^i$),
it follows that 
\begin{eqnarray*}
\EEW_{0,i}\le[\te{-\int_0^{\tau_t^+}h(Y_s)\td s}I_{\{Y^{+}_t=j, \tau^+_t<\infty, \tau_t^+<\rho\}}\ri] &=& I_{\{i= j\}}
\EEW\le[\te{-\tau_t^i h(i)}I_{\{\tau_t^i<\mbf e_i\}}\ri] = I_{\{i= j\}}\EEW\le[\te{-\tau_t^i (h_i+q_i)}\ri]\\
&=& I_{\{i= j, i\in E^+\}}\exp\le(-t(h_i + q_i)\ri)\\
&+& I_{\{i=j, i\in E^0\}}\exp\le(-t \sigma_i^{-2}\le(-\mu_i + \sqrt{\mu_i^2 + 2 (q_i+h_i)\sigma^2_i}\ri)\ri)\\
&=& I_{\{i= j\}}[1-\omega_i^+ t + o(t)]\qquad \text{as $t\downarrow 0$},
\end{eqnarray*}
where 
$\omega_i^+$
was defined in Lemma~\ref{lem:complex}
and
we used the fact that the Laplace transform of $\tau_a^i$ is given by
$$\WT{\mathbb{E}}[\te{-q \tau^i_a}] = \exp\le(-a\sigma_i^{-2}\le(-\mu_i + \sqrt{\mu_i^2 + 2q \sigma^2_i}\ri) \ri) \qquad a>0,\, q\in\mathbb C_{>0}.
$$

For every 
$h\in\mathcal H$
we can define the following matrices
\begin{eqnarray}
\label{eq:P_plus_minus}
\WT P_t^{h+}(i,j):=
\EEW_{0,i}\le[
\te{-\int_0^{\t_t^+}h(Y_s)\td s}I_{\{ Y_t^+ = j,
\t_t^+<\infty\}}\ri]\quad \text{for $i,j\in E^+\cup E^0$.}  
\end{eqnarray}
Note that definiton in~\eqref{eq:R+} 
implies 
that the identity
$
R_h^+(u)(i,j) = \int_0^\infty \te{-uy}\WT P^{h+}_{y}(i,j)\td y
$
holds for all
$u\in\CC_{> 0}$.
For every 
$h\in\mathcal H$,
$x\in(-\infty,t]$
and
$m\in E$
we have the follwing identity
\begin{eqnarray}
\nonumber
\EEW_{x,m}\le[
\te{-\int_0^{\t_t^+}h(Y_s)\td s}I_{\{ Y_t^+ = j,
\t_t^+<\infty\}}\ri] & = & 
\EEW_{0,m}\le[
\te{-\int_0^{\t_{t-x}^+}h(Y_s)\td s}I_{\{ Y_{t-x}^+ = j,
\t_{t-x}^+<\infty\}}\ri] \\
& = &  
\left(\eta_h^+ \WT P_{t-x}^{h+}\right)(m,j)
\quad \text{for $j\in E^+\cup E^0$,}  
\label{eq:SpaceInvarianceCond}
\end{eqnarray}
where
$\eta_h^+$
is defined in~\eqref{eq:etah+}. 
The first equality in~\eqref{eq:SpaceInvarianceCond}
is a consequence of the spacial homogeneity 
of the process
$\xi$.

Let 
$\ovl\xi_u=\sup_{s\leq u}\xi_s$ denote the running supremum of the 
stochastic process 
$\xi$.
The strong Markov property 
applied at the first jump time 
$\rho$
of the chain
$Y$
and~\eqref{eq:SpaceInvarianceCond}
imply that
\begin{eqnarray}
\nonumber
\EEW_{0,i}\le[\te{-\int_0^{\tau_t^+}h(Y_s)\td s}I_{\{Y^{+}_t=j, \tau^+_t<\infty, \tau_t^+\ge\rho\}}\ri] &=& 
\EEW_{0,i}\le[\te{-h(i)\rho}I_{\{\ovl \xi_{\rho}\leq t\}} \left(\eta_h^+ \WT P^{h+}_{t-\xi_\rho}\right)(Y_\rho,j)
\ri]\\
& = & \sum_{m\in E\backslash\{i\}} \frac{Q_0(i,m)}{q_i}
\EEW_{0,i}\le[\te{-h(i)\rho}I_{\{\ovl \xi_{\rho}\leq t\}} \left(\eta_h^+ \WT P^{h+}_{t-\xi_\rho}\right)(m,j)\ri].
\label{eq:FinalHopefully}
\end{eqnarray}
The last equality is a consequence of the fact that 
$Y_\rho$
is independent of the random vector
$(\rho,\xi_\rho,\ovl\xi_\rho)$
and
takes values in the set
$E\backslash\{i\}$
with 
$\widetilde{\PP}_{0,i}(Y_\rho=m)=Q_0(i,m)/q_i$.
Since 
$\rho$
is the first jump time of the chain
$Y$,
the vector 
$(\rho, \xi_\rho,\ovl\xi_\rho)$
has the same distribution as the vector
$(\mathbf e_i, X_{\mathbf e_i}^i,\ovl X_{\mathbf e_i}^i)$,
where as above
$\mathbf e_i$
is an exponential random variable with mean
$1/q_i$,
independent of the Brownian motion with drift
$X^i_t=\mu_i t+\sigma_i W_t$.
The symbol
$\ovl X_{\mathbf e_i}^i$
denotes the maximum of 
$X^i$
at the independent exponential time
$\mathbf e_i$.
Note that if
$i\in E^+$,
then 
$\ovl X_{t}^i=X_{t}^i=t $
for all
$t\in\RR_+$
and the expectation in~\eqref{eq:FinalHopefully}
is very easy to compute.

Assume now that 
$i\in E^0$.
Then the Wiener-Hopf factorisation implies that the random variables 
$\ovl X_{\mathbf e_u}^i$
and
$\ovl X_{\mathbf e_u}^i-X_{\mathbf e_u}^i$
are exponentially distributed  
with parameters
$$\sigma_i^{-2}\le(-\mu_i + \sqrt{\mu_i^2 + 2 u \sigma^2_i}\ri)\quad\text{and}\quad
\sigma_i^{-2}\le(\mu_i + \sqrt{\mu_i^2 + 2 u \sigma^2_i}\ri)$$
respectively
and independent
for any exponential random variable 
$\mathbf e_u$
with parameter 
$u>0$,
which is independent of
$X^i$. 
Therefore
the joint density 
$f_{\ovl X_{t}^i,\ovl X_{t}^i-X_{t}^i}$
satisfies the following identity for any
$x,y\in(0,\infty)$
\begin{eqnarray}
\label{eq:ProductOfIntegrals}
\int_0^\infty u\te{-ut}f_{\ovl X_{t}^i,\ovl X_{t}^i-X_{t}^i}(x,y)\td t =  \int_0^\infty u\te{-ut}f_{\ovl X_{t}}(x)\td t\>\>
\int_0^\infty u\te{-ut}f_{\ovl X_{t}^i-X_{t}^i}(y)\td t\quad \text{for all}\quad u>0,
\end{eqnarray}
where
$f_{\ovl X_{t}^i-X_{t}^i}$
and
$f_{\ovl X_{t}^i}$
are the densities of the corresponding random variables.
It is clear that there exists a unique extension 
to the complex half-plane
$\CC_{>0}$
of both sides of the formula
in~\eqref{eq:ProductOfIntegrals}
and that the following formulae hold
\begin{eqnarray}
\label{eq:Formula+}
\int_0^\infty \te{-ut}f_{\ovl X_{t}}(x)\td t & = &  \frac{\sigma_i^{-2}\le(-\mu_i + \sqrt{\mu_i^2 + 2 u \sigma^2_i}\ri)}{u}\te{-x\sigma_i^{-2}\le(-\mu_i + \sqrt{\mu_i^2 + 2 u \sigma^2_i}\ri)}
\qquad x>0, u\in\mbb C_{>0},\\
\int_0^\infty \te{-ut}f_{\ovl X_{t}^i-X_{t}^i}(y)\td t & = &  \frac{\sigma_i^{-2}\le(\mu_i + \sqrt{\mu_i^2 + 2 u \sigma^2_i}\ri)}{u}\te{-y\sigma_i^{-2}\le(\mu_i + \sqrt{\mu_i^2 + 2 u \sigma^2_i}\ri)}
\qquad y>0, u\in\mbb C_{>0}.
\label{eq:Formula-}
\end{eqnarray}
By substituting the identity~\eqref{eq:ProductOfIntegrals}
into the expectation~\eqref{eq:FinalHopefully},
applying the formulae in~\eqref{eq:Formula+} and~\eqref{eq:Formula-}
and taking the limit as 
$t$
tends to zero
we see that
the limits in \eqref{eq:Qh+ij} and \eqref{eq:Qh+ii} are as stated
for any
$h\in\mathcal H$.
Furthermore the formula in~\eqref{eq:Qhij+} holds.

The strong Markov property of $(\xi,Y)$ next implies that for any 
$h\in\mathcal H$ and any $t>0$ the matrices 
$\WT P^{h+}_t=(\WT P^{h+}_t(i,j)_t, i,j\in E^0\cup E^+)$ 
with  
$\WT P_t^{h+}(i,j)$ 
defined by~\eqref{eq:P_plus_minus} 
satisfy the system of 
ordinary differential equations
\begin{equation*}
\frac{\td}{\td t} \WT P^{h+}_t = \WT P^{h+}_t Q^{+}_h, \quad t>0, \qquad P^{h+}_0 = \mathbb I,
\end{equation*}
where $\mathbb I$ denotes an $N^+\times N^+$ 
identity matrix, the unique solution of which is given by 
$$\WT P^{h+}_t = \exp(Q^{+}_h t).$$ This proves that the matrix
$Q^{+}_h$ identified above satisfies \eqref{eq:Qh+}. Since 
$\WT P_t^{h+}(i,j)\to 0$ for all $i,j\in E^0\cup E^+$ as $t\to\infty$,  
it follows 
that all eigenvalues of $Q_h^+$ must have non-positive real part and therefore
$Q_h^+\in\mathbb D(N^+)$.
The proof of the existence of a matrix $Q^{-}_h\in\mathbb D(N^-)$ 
satisfying 
\eqref{eq:Qh-} is similar and omitted. \exit


\subsection{Proof of Theorem~\ref{thm:WienerHopf}}
%
(Existence) For any $h\in\mathcal H$ and
$x,\ell\in\RR$ 
define the matrices $\Phi^\pm_\ell(x)$ by
\begin{equation}\label{eq:phi+-}
\Phi^+_\ell(x) = \eta^+_h\exp\le(Q^+_h(\ell-x)\ri)\quad \quad
\Phi^-_\ell(x) = \eta^-_h\exp\le(Q^-_h(x-\ell)\ri),
\end{equation}
The proof of existence
rests on the martingale property of $M^+ = \{M^+_t\}_{t\ge 0}$ and $M^-=\{M^-_t\}_{t\ge 0}$ given by
\begin{equation}\label{eq:M+-}
M^+_t = \te{-\int_0^{t\wedge\t_\ell^+}h(Y_s)\td s} f_+\le(Y_{t\wedge\t_\ell^+},\xi_{t\wedge\t^+_\ell}\ri)\quad
\text{and}\quad M^-_t =
\te{-\int_0^{t\wedge\t_\ell^-}h(Y_s)\td s}f_-\le(Y_{t\wedge\t^-_\ell},\xi_{t\wedge\t^-_\ell}\ri)
\end{equation}
with
\begin{equation}\label{eq:f+-}
f_+(i,x) := e_i'\Phi^+_\ell(x)k_+,\quad\quad f_-(i,x) :=
e_i'\Phi^-_\ell(x)k_-,
\end{equation}
where $k_+$ and $k_-$ are $N^+-$ and $N^--$ column vectors,
respectively. 

The martingale property of $M^+$ follows from the equality
$$
M^+_t = \EEW_{x,i}\le[\te{-\int_0^{\tau^+_\ell}h(Y_s)\td s}
k_+\le(Y^+_\ell\ri)I_{\le\{\t^+_\ell < \infty\ri\}} |\mc G_t\ri]
$$
where $\{\mc G_t\}_{t\ge 0}$ denotes the filtration generated by $(\xi,Y)$.
To verify this identity observe first that the Markov property of $(\xi,Y)$ yields that
\begin{eqnarray*}
\lefteqn{\EEW_{x,i}\le[\te{-\int_0^{\tau^+_\ell}h(Y_s)\td s}
k_+\le(Y^+_\ell\ri)I_{\le\{\t^+_\ell < \infty\ri\}} |\mc G_t\ri]} \\ 
&=& 
\le.\te{-\int_0^{{t\wedge\t_\ell^+}}h(Y_s)\td s}
\EEW_{x,i}\le[\te{-\int_0^{\tau^+_\ell}h(Y_s)\td s}
k_+\le(Y^+_\ell\ri)I_{\le\{\t^+_\ell < \infty\ri\}}\ri]\ri|_{(x,i) = \le(\xi_{t\wedge\t^+_\ell}, Y_{t\wedge\t_\ell^+}\ri)}.
\end{eqnarray*}
Further, in view of  
the strong Markov property and 
spatial homogeneity of $\xi$, the expectation on the right-hand side of the previous display  
is for $x\leq \ell$ given by 
\begin{eqnarray}
\nn \lefteqn{\EEW_{x,i}\le[\te{-\int_0^{\tau^+_\ell}h(Y_s)\td s}
k_+\le(Y^+_\ell\ri)I_{\le\{\t^+_\ell < \infty\ri\}}\ri]}\\
\nn &=& \EEW_{0,i}\le[\te{-\int_0^{\tau^+_{\ell-x}}h(Y_s)\td s}k_+\le(Y^+_{\ell-x}\ri)
I_{\le\{\t^+_{\ell-x} < \infty\ri\}}\ri]\\
\nn &=& \sum_{j\in E^0\cup E^+} \EEW_{0,i}\le[\te{-\int_0^{\tau^+_{0}}h(Y_s)\td s}I_{\le\{Y_0^+ = j, \t^+_{0} < \infty\ri\}}\ri]
\EEW\le[\te{-\int_0^{\tau^+_{\ell-x}}h(Y_s)\td s}k_+\le(Y^+_{\ell-x}\ri)
I_{\le\{\t^+_{\ell-x} < \infty\ri\}}|Y_0=j,\xi_0=0\ri]\\
&=& e_i'\eta^+_h\exp\le(Q^+_h(\ell-x)\ri)  = e_i'\Phi^+_\ell(x) k_+ = 
f_+(i,x) \label{eq:ffexp}
\end{eqnarray}
where the last line follows by   
the definitions \eqref{eq:Qh+} and \eqref{eq:etah+} 
of $Q_h^+$ and $\eta_h^+$.

As $M^+$ is a martingale, an application of It\^o's lemma shows that $f_+ =
(f_+(i,u),i\in E)$ satisfies for all $u<\ell$
\begin{equation}
\label{ITOPHASE2} \mbox{$\frac{1}{2}$}s(i)^2 f''_+(i,u) + m (i)
f'_+(i,u) + \sum_{j}q_{ij}(f_+(j,u)-f_+(i,u)) =  0,
\end{equation}
where $ f'_+$ and $ f''_+$  denote the first and second derivatives of
$f_+$ with respect to $u$.
By substituting the expressions (\ref{eq:phi+-}) -- \eqref{eq:f+-}
into equation (\ref{ITOPHASE2}) we find, since $k^+$ was arbitrary,
that $Q^+_h$ and $\eta^+_h$ satisfy the first set of equations of the
system (\ref{systemnoiseWH3}). The proof for $Q^-_h$ and $\eta^-_h$
is analogous and omitted. 

(Uniqueness) Now we turn to the
proof of the uniqueness of the Wiener-Hopf factorization. To this
end, let $( W^+, G^+,  W^-,  G^-)$ be a complex matrix Wiener-Hopf factorization
and define the function $\WT f$ as 
$f_+$
in \eqref{eq:f+-}, but
replacing $ \eta^+$ and $ Q^+$ by $W^+$ and ${ G^+}$
respectively. Since the pair
$( W^+,  G^+)$ 
satisfies equation 
\eqref{systemnoiseWH3}, 
it follows by an application of
It\^{o}'s lemma, that 
$M'_t=\te{-\int_0^t h(Y_s)\td s}\WT f(Y_t, \xi_t)$ 
is a local martingale. In view of the facts that $G^+\in\mathbb D(N^+)$ 
and $h\in\mathcal H$ it follows that $M'$ is in fact 
bounded on $\{t\leq \t^+_\ell\}$. An application of Doob's optional
stopping theorem then yields that
\begin{align}\nn
\WT{f}(j,x) &= \EEW_{x,j}\le[\te{-\int_0^{t\wedge \t^+_\ell} h(Y_s)\td s} \WT f(Y_{t\wedge \t^+_\ell}, \xi_{t\wedge\t^+_\ell})\ri]\\
&= \EEW_{x,j}\le[\te{-\int_0^{\t^+_\ell} h(Y_s)\td s} \WT f(Y^+_{\ell},
\xi_{\t^+_\ell})I_{\{\t^+_\ell<\i\}}\ri] + \lim_{t\to\i}\EEW_{x,j}\le[\te{-\int_0^t h(Y_s)\td s}\WT f(Y_{t}, \xi_{t})I_{\{\t^+_\ell=\i\}}\ri]. \label{optionals}
\end{align}
By the definition of $\WT f$, the absence of positive jumps of
$\xi$ and \eqref{eq:ffexp}, the first expectation 
in (\ref{optionals}) is equal to
$$
\EEW_{x,j}\le[\te{-\int_0^{\t^+_\ell} h(Y_s)\td s}\WT f(Y^+_{\ell},\ell)I_{\{\t^+_\ell<\i\}}\ri] = \EEW_{x,j}\le[\te{-\int_0^{\t^+_\ell} h(Y_s)\td s}k_+(Y^+_{\ell})I_{\{\t^+_\ell<\i\}}\ri] 
= f_+(j,x)
$$
for $x\leq \ell$. The second term in (\ref{optionals}) is zero, since 
$\int_0^t I_{\{Y_s\in E_0\}}\td s\to\infty$ $\WT{\PP}_{x,j}$ almost surely 
on the event
$\{\tau_\ell^+=\infty\}$
for all $j\in E$ and $x\in\mathbb R$,  and $\min_{i\in E_0}\Re(h(i))>0$.
Thus $f_+=\WT f$ for all $N^+$-column vectors $k_+$ and we deduce 
that $ G^+= Q^+_h$ and $
W^+=\eta^+_h$. Similarly, one can show that $ G^-= Q^-_h$ and $
W^-=\eta^-_h$ and the uniqueness is proved. \exit

\subsection{Proof of Theorem \ref{thm:MainEmbed}}
Applying the strong Markov property at $\bar\tau$ and noting that 
$\bar\tau \leq \tau^\xi$ yields, in view of the representation 
\eqref{eq:Hxij} that
\begin{eqnarray*}
H^{x,i}_j(h,u)  &=&  
\WT{\EE}_{x,i}\left[\te{- \int_0^{\bar\tau} h(Y_s)
\td s}I_{\{Y_{\bar\tau}=j\}}F(\xi_{\bar\tau}, Y_{{\bar\tau}})\ri]\\
&=& \WT{\EE}_{x,i}\left[\te{- \int_0^{\bar\tau} h(Y_s)
\td s}I_{\{Y_{\bar\tau}=j, \tau^+_\ub<\tau_{\lb}^-\}}F(\ub, Y_{{\bar\tau}})\ri]
+ \WT{\EE}_{x,i}\left[\te{- \int_0^{\bar\tau} h(Y_s)
\td s}I_{\{Y_{\bar\tau}=j, \tau^+_\ub>\tau_{\lb}^-\}}F(\lb, Y_{{\bar\tau}})\ri]
\end{eqnarray*}
where 
$$
F(x,i) = \EEW_{x,i}[\te{- \int_0^{\tau^\xi} \WT h(Y_s)
\td s + u\xi_{\tau^\xi}}I_{\{Y_{\tau^\xi}=j\}}]
= \EEW_{x,i}[\te{u\xi_{\tau^\xi}}I_{\{Y_{\tau^\xi}=j\}}]
$$
since $\WT h(i) = h(i)I_{\{i\in E^0\}}$, by definition of $\WT h$.  
From the definition of $\tau^\xi$ it is straightforward to check that 
$$
\WT{\PP}_{x,i}[\tau^\xi=0]=1 \q\text{for $i\in E^0$ and $x\in\{\lb,\ub\}$},
$$
so that $F(\ub,i) = \te{u\ub}\delta_{ij}$ and $F(\lb,i) = \te{u\lb}\delta_{ij}$
if $i\in E^0$, where $\delta_{ij}$ denotes the Kronecker delta. 

Moreover, in view of the form \eqref{eq:Q0}--\eqref{eq:SiV} of $Q_h$ and 
the definition of phase-type distribution, 
it is clear that, conditionally on $Y_0=i\in E^+_j$ and $\xi_{0}=\ub$, 
$Y_{\tau^\xi}=j$ and $\tau^\xi\sim$ PH$(\delta_i,B^+_j)$, where 
$\delta_i$ is the vector with elements $\delta_i=(\delta_{ik})$.  
Therefore we find, using \eqref{prop:Phase-typeExpMom}, that for $u$ with 
$\Re(u)<\alpha_j^+$ 
and 
$i\in E^+_j$
\begin{eqnarray*}
F(\ub,i) &=& \EEW_{\ub,i}[\te{u\xi_{\tau^\xi}}]\\
&=& \te{u\ub}\WT{\EE}_{\ub,i}[\te{u{\tau^\xi}}]
= \te{u\ub}\le[(-u I^+_j - B^+_j)^{-1}(-B_j^+)\1\ri](i).
\end{eqnarray*} 
Similarly, it follows that for $i\in E_j^-$ and $u$ with 
$\Re(u)>-\alpha_j^-$
$$
F(\lb,i) = \te{u\lb}\le[(u I^-_j - B^-_j)^{-1}(-B_j^-)\1\ri](i).
$$ 







\subsection*{Acknowledgment} 
Research supported by EPSRC grant {EP/D039053}.
Most of the work was carried out while the first author was at
the Department of Mathematics of Imperial College London.
We thank an anonymous referee for the useful suggestions 
that improved the presentation of the paper.

\bibliographystyle{plain}
\bibliography{cite}


\end{document}